\def\ls{\lower4pt\hbox{${\buildrel < \over \sim}$}}
\def\gs{\lower4pt\hbox{${\buildrel > \over \sim}$}}

\documentclass[12pt, preprint]{aastex}

\slugcomment{Accepted for publication in {\it The Astrophysical Journal}}

\shorttitle{Multiwavelength Observations of BL~Lac in 2000}
\shortauthors{B\"ottcher et al.}

\begin{document}

\title{Coordinated Multiwavelength Observations of BL~Lacertae in 2000}

\author{M. B\"ottcher\altaffilmark{1}, 
A. P. Marscher\altaffilmark{2}, M. Ravasio\altaffilmark{3}, M. Villata\altaffilmark{4}, 
C. M. Raiteri\altaffilmark{4}, H. D. Aller\altaffilmark{5}, M. F. Aller\altaffilmark{5}, 
H. Ter\"asranta\altaffilmark{6}, O. Mang\altaffilmark{7}, G. Tagliaferri\altaffilmark{3},
F. Aharonian\altaffilmark{8},
H. Krawczynski\altaffilmark{9},
O.~M.~Kurtanidze \altaffilmark{10,11,12},
M.~G.~Nikolashvili \altaffilmark{10},
M.~A.~Ibrahimov \altaffilmark{13,14},
I.~E.~Papadakis \altaffilmark{15,16},
K.~Tsinganos \altaffilmark{15},
K.~Sadakane \altaffilmark{17},
N.~Okada \altaffilmark{17},
L.~O.~Takalo \altaffilmark{18},
A.~Sillanp\"a\"a \altaffilmark{18},
G.~Tosti \altaffilmark{19},
S.~Ciprini \altaffilmark{19},
A.~Frasca \altaffilmark{20},
E.~Marilli \altaffilmark{20},
R.~Robb \altaffilmark{21},
J.~C.~Noble \altaffilmark{22},
S.~G.~Jorstad \altaffilmark{22},
V.~A.~Hagen-Thorn \altaffilmark{35},
V.~M.~Larionov \altaffilmark{23},
R.~Nesci \altaffilmark{24},
M.~Maesano \altaffilmark{24},
R.~D.~Schwartz \altaffilmark{25},
J.~Basler \altaffilmark{25},
P.~W.~Gorham \altaffilmark{26},
H.~Iwamatsu \altaffilmark{27},
T.~Kato \altaffilmark{27},
C.~Pullen \altaffilmark{28},
E.~Ben\'{\i}tez \altaffilmark{29},
J.~A.~de~Diego \altaffilmark{29},
M.~Moilanen \altaffilmark{30},
A.~Oksanen \altaffilmark{30},
D.~Rodriguez \altaffilmark{31},
A.~C.~Sadun \altaffilmark{32},
M.~Kelly \altaffilmark{32},
M.~T.~Carini \altaffilmark{33},
H.~R.~Miller \altaffilmark{34},
S.~Catalano \altaffilmark{20}
D.~Dultzin-Hacyan \altaffilmark{29},
J.~H.~Fan \altaffilmark{35},
G.~Ghisellini \altaffilmark{3},
R.~Ishioka \altaffilmark{27},
H.~Karttunen \altaffilmark{18},
P.~Kein\"anen \altaffilmark{18},
N.~A.~Kudryavtseva \altaffilmark{23},
M.~Lainela \altaffilmark{18},
L.~Lanteri \altaffilmark{4},
E.~G.~Larionova \altaffilmark{23},
K.~Matsumoto \altaffilmark{27},
J.~R.~Mattox \altaffilmark{36},
I.~McHardy \altaffilmark{38},
F.~Montagni \altaffilmark{24},
G.~Nucciarelli \altaffilmark{19},
L.~Ostorero \altaffilmark{37},
J.~Papamastorakis \altaffilmark{15,16},
M.~Pasanen \altaffilmark{18},
G.~Sobrito \altaffilmark{4}, and
M.~Uemura \altaffilmark{27}
}

\altaffiltext{1}{Department of Physics and Astronomy, Clippinger 339, Ohio University, \\
Athens, OH 45701, USA}
\altaffiltext{2}{Department of Astronomy, Boston University, 725 Commonwealth Ave.,\\
Boston, MA 02215, USA}
\altaffiltext{3}{Osservatorio Astronomico di Brera, Via Bianchi 46, I-23807 Merate, Italy}
\altaffiltext{4}{Istituto Nazionale di Astrofisica (INAF), Osservatorio Astronomico di Torino,\\
Via Osservatorio 20, I-10025 Pino Torinese, Italy}
\altaffiltext{5}{Department of Astronomy, University of Michigan, 810 Dennison Building, \\
Ann Arbor, MI 48109-1090, USA}
\altaffiltext{6}{Mets\"ahovi Radio Observatory, Helsinki University of Technology, \\
Mets\"ahovintie 114, 02540 Kylm\"al\"a, Finland}
\altaffiltext{7}{Institut f\"ur Experimentelle und Angewandte Physik, Universit\"at Kiel,\\
Leibnitzstra\ss e 15 -- 19, D-24118 Kiel, Germany}
\altaffiltext{8}{Max-Planck-Institut f\"ur Kernphysik, Postfach 10 39 80, \\
D-69029 Heidelberg, Germany}
\altaffiltext{9}{Physics Department, Washington University, 1 Brookings Drive CB 1105, \\
St. Louis, MO 63130, USA}
\altaffiltext{10}{Abastumani Observatory, 383762 Abastumani, Georgia}
\altaffiltext{11}{Astrophysikalisches Institute Potsdam, An der Sternwarte 16, \\
D- 14482 Potsdam, Germany}
\altaffiltext{12}{Landessternwarte Heidelberg-K\"onigstuhl, K\"onigstuhl 12, \\
D-69117 Heidelberg, Germany}
\altaffiltext{13}{Ulugh Beg Astronomical Institute, Academy of Sciences of Uzbekistan, \\
33 Astronomical Str., Tashkent 700052, Uzbekistan}
\altaffiltext{14}{Isaac Newton Institute of Chile, Uzbekistan Branch}
\altaffiltext{15}{Physics Department, University of Crete, 710 03 Heraklion, Crete, Greece}
\altaffiltext{16}{IESL, Foundation for Research and Technology-Hellas, \\
711 10 Heraklion, Crete, Greece}
\altaffiltext{17}{Astronomical Institute, Osaka Kyoiku University, Kashiwara-shi, \\
Osaka, 582-8582 Japan}
\altaffiltext{18}{Tuorla Observatory, 21500 Piikki\"o, Finland}
\altaffiltext{19}{Osservatorio Astronomico, Universit\`a di Perugia, Via B.\ Bonfigli, \\
I-06126 Perugia, Italy}
\altaffiltext{20}{Osservatorio Astrofisico di Catania, Viale A.\ Doria 6, \\
I-95125 Catania, Italy}
\altaffiltext{21}{Department of Physics and Astronomy, University of Victoria, \\
British Columbia, Canada}
\altaffiltext{22}{Institute for Astrophysical Research, Boston University, \\
725 Commonwealth Ave., Boston, MA 02215, USA}
\altaffiltext{23}{Astronomical Institute, St.-Petersburg State University, \\
Bibliotechnaya Pl.\ 2, Petrodvoretz, 198504 St.-Petersburg, Russia}
\altaffiltext{24}{Dipartimento di Fisica, Universit\`a La Sapienza, Piazzale A.\ Moro 2, \\
I-00185 Roma, Italy}
\altaffiltext{25}{Department of Physics and Astronomy, University of Missouri-St.\ Louis, \\
8001 Natural Bridge Road, St.\ Louis, MO 63121, USA}
\altaffiltext{26}{Jet Propulsion Laboratory, California Institute of Technology, \\
4800 Oak Grove Drive, Pasadena, CA 91109, USA}
\altaffiltext{27}{Department of Astronomy, Faculty of Science, Kyoto University, \\
Kyoto, Japan}
\altaffiltext{28}{Clarke and Coyote Astrophysical Observatory, PO Box 930, \\
Wilton, CA 95693, USA}
\altaffiltext{29}{Instituto de Astronom\'{\i}a, UNAM, Apdo.\ Postal 70-264, \\
04510 M\'exico DF, Mexico }
\altaffiltext{30}{Nyr\"ol\"a Observatory, Jyv\"askyl\"an Sirius ry, Kyllikinkatu 1, \\
40950 Jyv\"askyl\"a, Finland}
\altaffiltext{31}{Guadarrama Observatory, C/ San Pablo 5, Villalba 28409, Madrid, Spain}
\altaffiltext{32}{Department of Physics, University of Colorado at Denver, PO Box 173364, \\
Denver, CO 80217-3364, USA}
\altaffiltext{33}{Department of Physics and Astronomy, Western Kentucky University, \\
1 Big Red Way, Bowling Green, KY 42104, USA}
\altaffiltext{34}{Department of Physics and Astronomy, Georgia State University, \\
Atlanta, GA 30303, USA}
\altaffiltext{35}{Center for Astrophysics, Guangzhou University, Guangzhou 510400, China}
\altaffiltext{36}{Department of Chemistry, Physics, \& Astronomy, Francis Marion University, \\
PO Box 100547, Florence, SC 29501-0547, USA}
\altaffiltext{37}{Dipartimento di Fisica Generale, Universit\`a di Torino, \\
Via P.\ Giuria 1, I-10125 Torino, Italy}
\altaffiltext{38}{Department of Physics and Astronomy, University of Southampton,
Highfield, Southampton, SO17 1BJ, Great Britain}

\begin{abstract}
BL~Lacertae was the target of an extensive multiwavelength
monitoring campaign in the second half of 2000. Simultaneous or 
quasi-simultaneous observations were taken at radio (UMRAO and 
Mets\"ahovi) and optical(WEBT collaboration) frequencies, in 
X-rays (BeppoSAX and RXTE), and at VHE gamma-rays (HEGRA). 
The WEBT optical campaign achieved an unprecedented time 
coverage, virtually continuous over several 10 -- 20~hour
segments. It revealed intraday variability on time scales 
of $\sim 1.5$~hours and evidence for spectral hardening
associated with increasing optical flux. During the campaign,
BL~Lacertae underwent a major transition from a rather quiescent
state prior to September 2000, to a flaring state for the rest
of the year. This was also evident in the X-ray activity of the
source. BeppoSAX observations on July 26/27 revealed a rather 
low X-ray flux and a hard spectrum, while a BeppoSAX pointing
on Oct. 31 -- Nov. 2, 2000, indicated significant variability 
on time scales of $\lesssim$~a few hours, and provided evidence 
for the synchrotron spectrum extending out to $\sim 10$~keV during 
that time. During the July 26/27 observation, there is a tantalizing, 
though not statistically significant, indication of a time delay of 
$\sim 4$ -- 5~hr between the {\it Beppo}SAX and the R-band light 
curve. Also, a low-significance detection of a time delay of 15~d
between the 14.5~GHz and the 22~GHz radio light curves is reported. 
Several independent methods to estimate the co-moving magnetic field 
in the source are presented, suggesting a value of $\sim 2 \, e_B^{2/7}$~G,
where $e_B$ is the  magnetic-field equipartition factor w.r.t. the 
electron energy density in the jet.
\end{abstract}


\section{Introduction}

BL Lacertae (= 1ES~2200+420; z = 0.069) was historically the prototype 
of the BL Lac class of active galactic nuclei (AGN). These objects are 
characterized by continuum properties similar to those of flat-spectrum 
radio quasars (non-thermal optical continuum, high degree of linear
polarization, rapid variability at all wavelengths, radio jets
with individual components often exhibiting apparent superluminal
motion), but do usually show only weak emission or absorption lines 
(with equivalent width in the rest-frame of the host galaxy of 
$< 5$~\AA), if any. In BL~Lacertae itself, however, H$\alpha$ 
(and H$\beta$) emission lines have been detected during a period of 
several weeks in 1995 \citep{ver95,cor96}, and in 1997 \citep{cor00}. 
Superluminal motion of $\beta_{\rm app}$ up to $(5.0 \pm 0.2) \, 
h^{-1} \approx (7.1 \pm 0.3)$ has been observed in this object 
\citep{denn00}.

BL Lac objects and flat spectrum radio quasars (FSRQs) are 
commonly unified in the AGN class of blazars. Sixty-five blazars 
have been detected and identified with high confidence in 
high energy ($> 100$~MeV) gamma-rays by the EGRET instrument 
on board the {\it Compton Gamma-Ray Observatory} \citep{hartman99,mhr01}. 
To date, 6 blazars have been detected at very high energies ($> 300$~GeV) 
with ground-based air \v Cerenkov detectors 
\citep{punch92,quinn96,catanese98,chadwick99,aharonian02,horan02,holder03}.
All of these belong to the sub-class of high-frequency peaked BL~Lac 
objects (HBLs). The field of extragalactic GeV -- TeV astronomy is currently 
one of the most rapidly expanding research areas in astrophysics. The 
steadily improving flux sensitivities of the new generation of air 
\v Cerenkov telescope arrays and their decreasing energy thresholds
\citep[for a recent review see, e.g.,][]{weekes02}, provides a growing 
potential to extend their extragalactic-source list towards intermediate 
and even low-frequency peaked BL~Lac objects (LBLs) with lower $\nu F_{\nu}$ 
peak frequencies in their broadband spectral energy distributions (SEDs). 
Detection of such objects at energies $\sim 40$ -- 100~GeV might provide 
an opportunity to probe the intrinsic high-energy cutoff of their SEDs 
since at those energies, $\gamma\gamma$ absorption due to the intergalactic 
infrared background is still expected to be negligible at redshifts of 
$z \lesssim 0.2$ \citep{djs02}. There has even been a claimed detection
of BL~Lacertae in 1998 with the \v Cerenkov telescope of the Crimean 
Astrophysical Observatory \citep{neshpor01}; however, this detection
could not be confirmed by any other group so far \citep[e.g.,][]{aharonian00}.

BL~Lacertae is classified as an LBL. From an interpolation between the
GHz radio spectrum and the IR - optical spectrum, it can be inferred
that its low-frequency spectral component typically peaks at mm to
$\mu$m wavelengths, while the high-frequency component seems to peak
in the multi-MeV -- GeV energy range. BL~Lacertae has been the target 
of many radio, optical, X-ray, and $\gamma$-ray observations in the past, 
and has been studied in detail during various intensive multiwavelength 
campaigns \citep[e.g.][]{bloom97,sambruna99,madejski99,ravasio02,villata03}. 
It is a particularly interesting object for detailed X-ray studies: this is
the region of the electromagnetic spectrum where the two broad components 
of the multiwavelength SEDs of BL~Lacertae (and other LBLs) are overlapping 
and intersecting. X-ray observations of this source at different epochs 
show significant flux and spectral variability, indicating that the
X-ray emission is at times dominated by the high-energy end of the 
synchrotron emission, while at other occasions it is dominated by the
low-frequency portion of the high-energy bump of the SED. In fact, BL~Lacertae 
has repeatedly shown a concave shape \citep[e.g.,][]{madejski99,ravasio02}, 
with rapid variability mainly restricted to the low-energy excess portion 
of the spectrum \citep[e.g.,][]{ravasio02,ravasio03}.

In the framework of relativistic jet models, the low-frequency (radio
-- optical/UV) emission from blazars is interpreted as synchrotron
emission from nonthermal electrons in a relativistic jet. The
high-frequency (X-ray -- $\gamma$-ray) emission could either be
produced via Compton upscattering of low frequency radiation by the
same electrons responsible for the synchrotron emission \citep[leptonic
jet models; for a recent review see, e.g.,][]{boettcher02}, or 
due to hadronic processes initiated by relativistic protons 
co-accelerated with the electrons \citep[hadronic models, for a recent
discussion see, e.g.,][]{muecke01,muecke03}. 

While simultaneous broadband spectra are very useful to constrain
blazar jet models, there still remain severe ambiguities in their
interpretation w.r.t. the dominant electron cooling, injection,
and acceleration mechanisms, as very drastically illustrated for
the case of W~Comae by \cite{bmr02}. Those authors have also
demonstrated that a combination of broadband spectra with timing 
and spectral variability information, in tandem with time-dependent
model simulations \citep[e.g.,][]{bc02,krawczynski02} can help to break 
some of these degeneracies. For this reason, we organized an intensive 
multiwavelength campaign to observe BL~Lacertae in the second half 
of 2000 at as many frequencies as possible, putting special 
emphasis on detailed variability information. In \S 
\ref{observations}, we describe the observations carried out 
during the campaign and present light curves in the various 
energy bands. The diverse spectral variability patterns are
discussed in \S \ref{variability}, and the results of 
our search for inter-band cross-correlations and
time lags are presented in \S \ref{crosscorrelations}. 
The source underwent a dramatic state transition from a rather
quiescent state until mid-September 2000, to a very active
state which lasted throughout the remainder of the year. We
have obtained two detailed simultaneous broadband spectra of
BL~Lacertae, one before and one after this transition. The
resulting SEDs are presented in \S \ref{spectra}. In \S
\ref{parameter_estimates} we use our results to derive 
estimates of generic model parameters, in particular of 
the co-moving magnetic field, independent of the details of 
any specific model. In a companion paper (B\"ottcher \&
Reimer 2003, in preparation), we will use leptonic and 
hadronic models used to fit the spectra and variability
patterns found in this campaign. We summarize in \S \ref{summary}.

Throughout this paper, we refer to $\alpha$ as the energy 
spectral index, $F_{\nu}$~[Jy]~$\propto \nu^{-\alpha}$. A 
cosmology with $\Omega_m = 0.3$ and $\Omega_{\Lambda} = 0.7$ 
and $H_0 = 70$~km~s$^{-1}$~Mpc$^{-1}$ is used.

\section{\label{observations}Observations, data reduction, 
and light curves}

BL~Lacertae was observed in a co-ordinated multiwavelength 
campaign at radio, optical, X-ray, and VHE $\gamma$-ray energies 
during the period mid-May 2000, until the end of the year. The
overall timeline of the campaign, along with the measured long-term
light curves at radio, optical, and X-ray frequencies, is illustrated
in Fig. \ref{timeline}. 

\subsection{\label{radio}Radio observations}

At radio frequencies, the object was monitored using the University 
of Michigan Radio Astronomy Observatory (UMRAO) 26~m telescope, 
at 4.8, 8, and 14.5~GHz, and the 14~m Mets\"ahovi Radio Telescope 
of Helsinki University of Technology, at 22 and 37~GHz. The radio 
light curves are shown in the top two panels of Fig. \ref{timeline}.
At the lower frequencies (4.8, 8, and 14.5~GHz), they show evidence 
for flux variability on a $\sim 30$~\% level on time scales of 
$\sim 1$~month and track each other closely. Superimposed on the
large-amplitude variability on a $\sim 1$~month time scale, the 
4.8-GHz and 8-GHz radio light curves exhibit low-amplitude variability
on time scales of a few days. The discrete autocorrelation functions
\citep{ec88} of these light curves indicate a sharp decline on a time 
scale of $\sim 4$~days. 

The higher-frequency radio light curves indicate more erratic variability, 
with flux variations of $\sim 25$~\% within a few days. However, those
variability patterns are clearly under-sampled in our data set, so that
a more detailed analysis might not be meaningful at this point.

\subsection{\label{optical}Optical observations}

Focusing on an originally planned core campaign period of July 17 -- 
Aug. 11, BL~Lacertae was the target of an intensive optical
campaign by the Whole Earth Blazar Telescope 
\citep[WEBT][see also 
{\tt http://www.to.astro.it/blazars/webt/}]{villata00,raiteri01}, 
in which 24 optical telescopes throughout the northern 
hemisphere participated. Details of the data collection,
analysis, cross-calibration of photometry from different 
observatories, etc. pertaining to the WEBT campaign have
been published in \cite{villata02}. 
Observations were made in the standard U, B, V, R, and 
I bands. For the purpose of broadband spectroscopy, the fluxes 
were corrected for extinction and reddening using a B-band 
extinction value of $A_B = 1.42$ \citep{schlegel98} and
the extinction law of \cite{cardelli89}. The contribution
of the host galaxy was subtracted as described in detail
in \cite{villata02}.

Fig. \ref{timeline} illustrates that BL~Lacertae was in a
rather quiescent state during the core campaign, in which
the densest light curve sampling was obtained. However, the
source underwent a dramatic state transition to an extended 
high state in mid-September 2000. For this reason, the WEBT 
campaign was extended until early January of 2001, although 
with less dense time coverage than during the core campaign. 

The WEBT campaign returned optical (R-band) light curves of
unprecedented time coverage and resolution. Fig. \ref{optical_lc_core}
shows the R-band light curves over the entire core campaign
\citep[see also Fig. 2 of][]{villata02}. The bottom panel of 
Fig. \ref{r_lecs_lc} \citep[see also Figs. 3 -- 5 of][]{villata02} 
illustrates the microvariability measured for two individual 
nights during this period. Brightness variations of $\Delta R 
\sim 0.35$, corresponding to flux variations of $(\Delta F)/F 
\sim 0.4$, within $\sim 1.5$~hr have been found. Such rapid 
microvariability is not exceptional for this source and had 
been observed before on several occasions
\citep[e.g.,][]{miller89,carini92,nesci98,sn98,cc01}. It is also 
confirmed by the autocorrelation function of the R-band light curve, 
which can be well fitted with an exponential with a decay time scale of
$\sim 2$~hr. The observed variability time scale places a constraint 
on the size of the emitting region of $R \lesssim 1.6 \times 10^{14} 
\, D$~cm, where $D = \left(\Gamma [ 1 - \beta\cos\theta_{\rm obs}] 
\right)^{-1}$ is the Doppler beaming factor. 

\subsection{\label{xrays}X-ray observations}

At X-ray energies, BL~Lacertae was observed with the {\it BeppoSAX}
Narrow Field Instruments (NFI) in the energy range 0.1 -- 200~keV
in two pointings on July 26 -- 27 and Oct. 31 -- Nov. 2, 2000
\citep{ravasio03}. In addition, the source was monitored by 
the {\it Rossi X-ray Timing Explorer (RXTE)} Proportional Counter
Array (PCA) in 3 short (individual exposures ranging from a few 
hundred to $\sim 2000$~s) pointings per week (Marscher et al. 2003, 
in preparation). Table \ref{x_obs} summarizes the precise times 
of the two {\it BeppoSAX} pointings and of the {\it RXTE PCA}
observations closest to the {\it BeppoSAX} ones, along with the
results of the spectral analysis. The details of the {\it BeppoSAX} 
observations and the data analysis methods have been published in 
\cite{ravasio03}. Note that the PCA had observed BL~Lacertae exactly 
simultaneously with the {\it BeppoSAX} pointing during the July 26 
-- 27 observation, while an observation on Nov. 2 started 1~hr after 
the second {\it BeppoSAX} pointing. 

\begin{figure}
\plotone{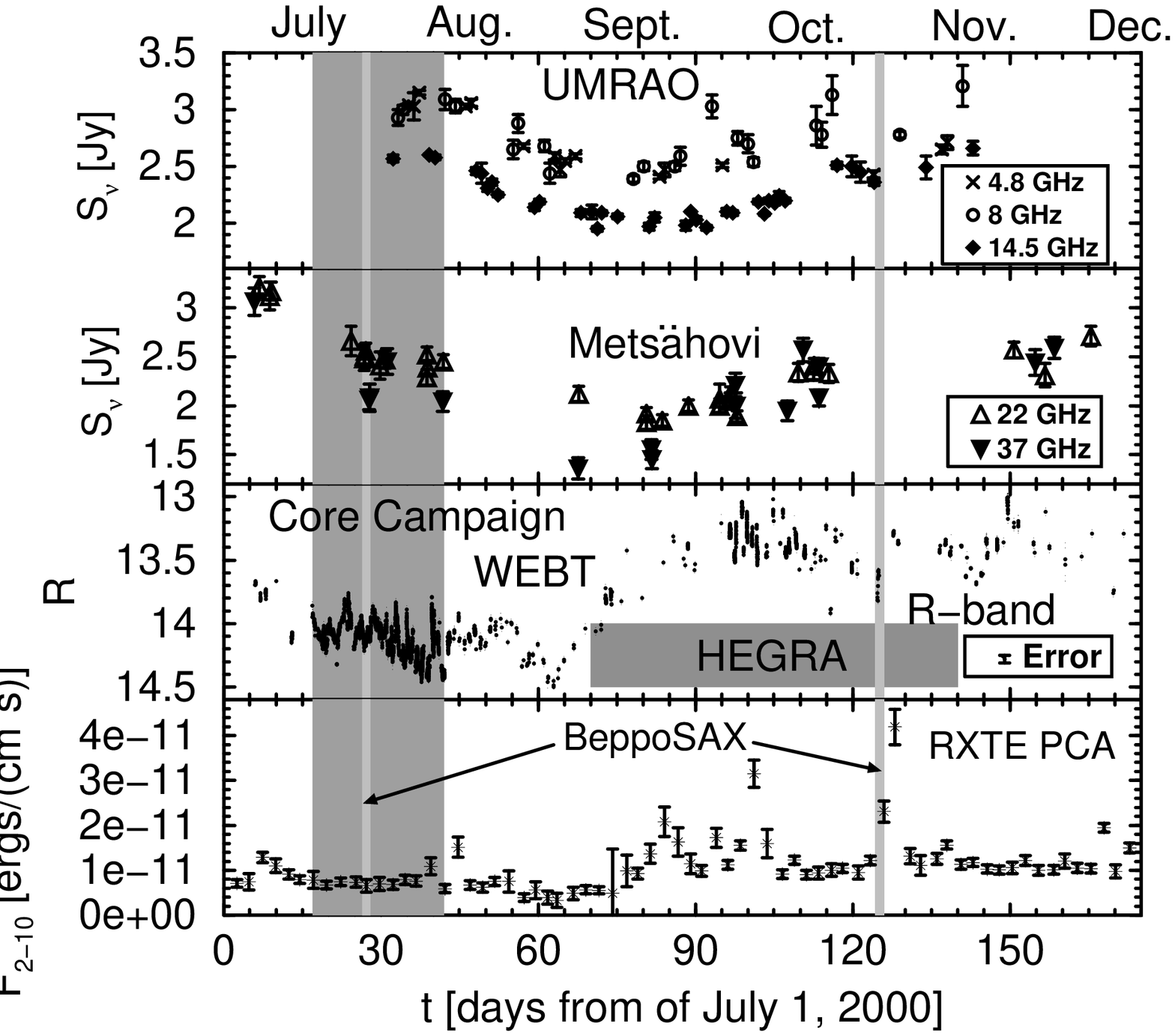}
\caption{Time line of the broadband campaign on BL Lacertae in 2000.
The dark-shaded areas indicate the duration of the core campaign,
July 17 -- August 11, 2000 and of the HEGRA observations (collecting a total
of 10.5 hours of on-source time); the light-shaded areas indicate the
times of our two BeppoSAX pointings. For clarity, the error bars on the
R-band magnitudes have not been plotted individiually. The inset in the
lower-right corner of the third panel shows the typical error bar for 
measurements outside the core campaign; the errors are typically a factor 
of 3 smaller than this during the core campaign.}
\label{timeline}
\end{figure}

\begin{figure}
\plotone{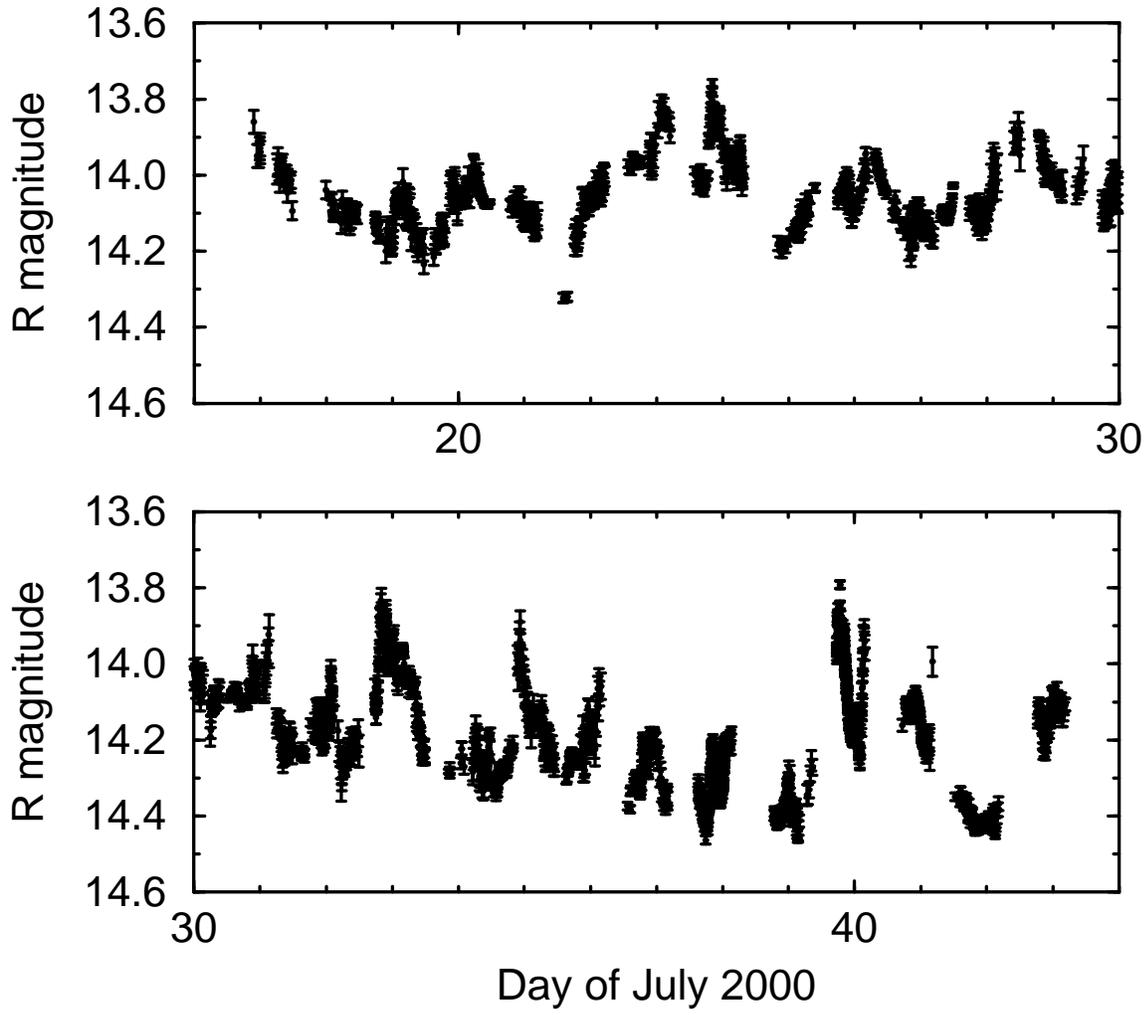}
\caption{Optical (R-band) light curve of BL~Lacertae during the core
campaign of July 17 -- August 11, 2000 \citep{villata02}.}
\label{optical_lc_core}
\end{figure}

The drastic change of the activity state of BL~Lacertae in 
mid-September observed in the optical range is accompanied by
several large flares in the PCA light curve over a $\sim 2$~months
period (see Fig. \ref{timeline}), but not by a similarly extended 
high flux state as seen in the optical. In fact, while the 
average flux level increased only slightly, a higher level of 
activity was indicated by a higher degree of variability: 
The average PCA 2 -- 10~keV flux increased from $7.1 \times 
10^{-12}$~ergs~cm$^{-2}$~s$^{-1}$ for the period MJD 51725 -- 
51795 (end of June -- early September 2000) to $1.2 \times 
10^{-11}$~ergs~cm$^{-2}$~s$^{-1}$ for the period MJD 51800 --
51910 (mid-September -- end of December 2000). A constant fit
to the quiescent phase resulted in a reduced $\chi_{\nu}^2
= 3.47$, which increased to $\chi_{\nu}^2 = 6.63$ for the
remainder of the year. This quantifies the drastically
increased 2 -- 10~keV X-ray variability on time scales of a 
few days probed by the PCA monitoring observations, in the 
active state. However, it also shows that BL~Lacertae exhibits 
significant X-ray variability on this time scale even in the 
quiescent state. 

Fig. \ref{timeline} also shows that we were extremely lucky 
to catch BL~Lacertae in an exceptional X-ray outburst during
our second {\it BeppoSAX} pointing. In fact, the 2 -- 10~keV
flux measured on Oct. 31 -- Nov. 2, 2000 was the highest ever
detected by {\it Beppo}SAX from this source. Interestingly, 
the R-band lightcurve indicates a relatively low optical flux, 
compared to the average flux level after mid-September 2000, 
coincident with this X-ray outburst. If the optical and soft 
X-ray fluxes are due to synchrotron emission from the same 
population of electrons, this could indicate a hardening of 
the electron spectrum during the flaring state. However, as
pointed out and discussed by \cite{ravasio03} and in \S \ref{spectra}, 
the optical and X-ray spectra during this observation can not 
be connected by a smooth power-law: the optical fluxes are
significantly below a power-law extrapolation of the {\it Beppo}SAX
LECS + MECS spectrum. This could possibly indicate that the
optical and X-ray fluxes are coming from separate regions 
along the jet, possibly also associated with substantial time
lags between these emissions.

\subsubsection{\label{july26}July 26 -- 27}

During the July 26 -- 27 {\it BeppoSAX} observation, the source
was in a low flux and activity state. This only allowed a rather
restricted spectral analysis and the extraction of meaningful
light curves with a binning of no less than 1~hr. For details
of the spectral and timing analysis, see \cite{ravasio03}. They 
tested several spectral models, including a single power-law
with free $N_H$, a single power-law with a fixed value of
$N_H$, and a broken power-law model. In the following, we will
concentrate on the results from the analysis with $N_H = 2.5 \times
10^{21}$~cm$^{-2}$, resulting from the dust-to-gas ratio suggested
by \cite{ryter96} with $A_B = 1.42$ and the de-reddening law as 
mentioned above in \S \ref{optical}, and consistent with previous
spectral analyses of X-ray observations of BL~Lacertae
\citep{sambruna99,madejski99,ravasio02}. The fit resulted in $\alpha
= 0.8 \pm 0.1$. This is perfectly consistent with the result from 
the contemporaneous {\it RXTE} PCA observation. The spectral
index of $\alpha = 0.8 \pm 0.1$ confirms the low-activity state of 
the source at that time and indicates that the entire X-ray spectrum 
might have been dominated by the low-frequency end of the high-energy 
component of the broadband SED of BL~Lacertae.

The short-term LECS ([0.7 -- 2]~keV) and MECS ([2 -- 10]~keV) 
lightcurves of BL~Lacertae during this observation \citep[see 
Fig. 3 of][]{ravasio03} display a large (factor $> 2$) flare
on a time scale of $\sim 4$~hr, while the source appears less 
variable at higher energies. This behavior has been noted in
this source before \citep[e.g.,][]{ravasio02}, and is even more 
obvious in the Oct 31 -- Nov. 2 observation (see next subsection). 
The low count rate and relatively short exposure time prevents a
more detailed analysis of variability features of this observation.

\subsubsection{\label{nov1}Oct. 31 -- Nov. 2}

During the second {\it BeppoSAX} pointing on Oct. 31 -- Nov. 2, 
2000, we measured the highest 2 -- 10~keV flux ever observed
with {\it Beppo}SAX from BL~Lacertae. The LECS + MECS spectrum 
in the 0.3 -- 10~keV range was well fitted with a power-law model 
with fixed $N_H = 2.5 \times 10^{21}$~cm$^{-2}$, revealing a steep 
X-ray spectrum with $\alpha = 1.56 \pm 0.03$ \citep{ravasio03}. 
As for the Nov. 26 -- 27 observation, the spectral fitting results
were consistent with the results from the PCA observations beginning
$\sim 1$~hr after the {\it Beppo}SAX pointing. In this observation, 
BL~Lacertae was also significantly detected by the PDS up to $\sim 50$~keV.
The PDS spectrum indicates a significant spectral hardening
beyond $\sim 10$~keV with a best-fit spectral index $\alpha_{\rm PDS}
= 0.56$ for which, however, no error could be estimated due to the
poor photon statistics \citep{ravasio03}. The soft shape of the
LECS + MECS spectrum clearly indicates that it was dominated by
the high-energy end of the low-energy (synchrotron) component in 
this observation. Evidence for the synchrotron component at soft 
X-ray energies had been found in BL~Lacertae before 
\citep{madejski99,ravasio02}, but this is the first time 
that this behavior was observed extending all the way out 
to $\sim 10$~keV. The spectral hardening evident in the PDS
spectrum might indicate the onset of the high-energy component
beyond $\sim 10$~keV.

\begin{figure}
\plotone{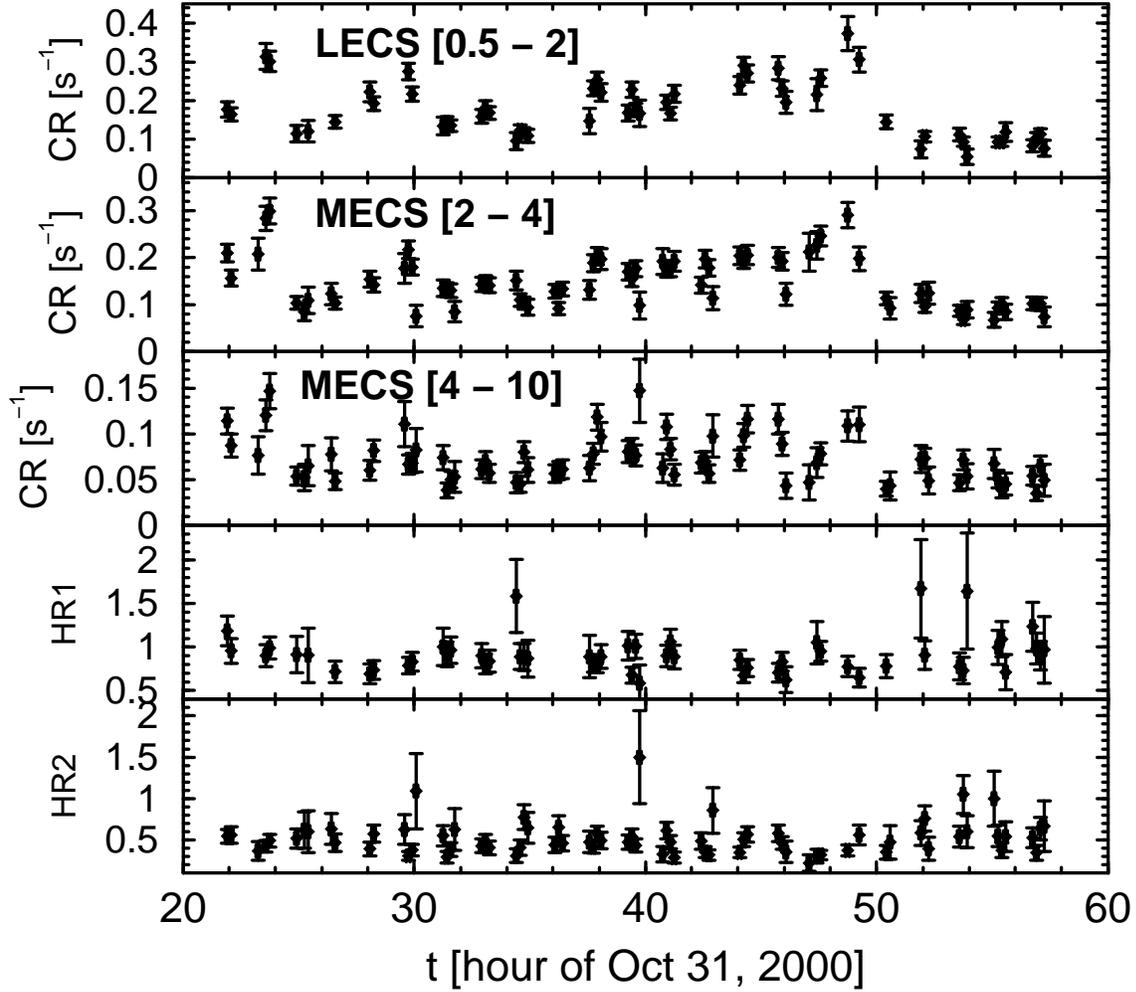}
\caption{Light curves in three different energy channels of the {\it BeppoSAX}
LECS + MECS observations on Oct. 31 -- Nov. 2, 2000. The two lower panels show
the hardness ratios HR1 = MECS [2 - 4] / LECS [0.5 - 2] and HR2 = MECS [4 - 10] 
/ MECS [2 - 4].}
\label{sax_lc}
\end{figure}

Fig. \ref{sax_lc} displays the LECS and MECS light curves in
three different energy channels during the second {\it BeppoSAX}
pointing, along with the two hardness ratios: HR1 = MECS [2 - 4]
/ LECS [0.5 - 2] and HR2 = MECS [4 - 10] / MECS [2 - 4]. The 
LECS and MECS light curves show significant variability in all
energy channels, with flux variations of factors of $\sim 3$ -- 
4 on time scales down to $\sim 1$ -- 2~hr. The PDS counts were 
consistent with no variability \citep[constancy probability of $\sim
96$~\%,][]{ravasio03}. \cite{ravasio03} have calculated the 
normalized excess variance parameter $\sigma_{\rm rms}^2$ for 
the three LECS and MECS energy channels and found that 
$\sigma_{\rm rms}^2$ is slightly decreasing with increasing 
photon energy. The LECS and MECS light curves show several 
individual, well-resolved flares, (e.g., at $\sim 29$~hr and 
$\sim 49$~hr, see Fig. \ref{sax_lc}). Those flares seem to
suggest slightly longer rise than decay time scales, but due 
to the limited photon statistics, a meaningful, more quantitative 
assessment of light curve asymmetries is not possible with our
present data. 

\begin{figure}
\plotone{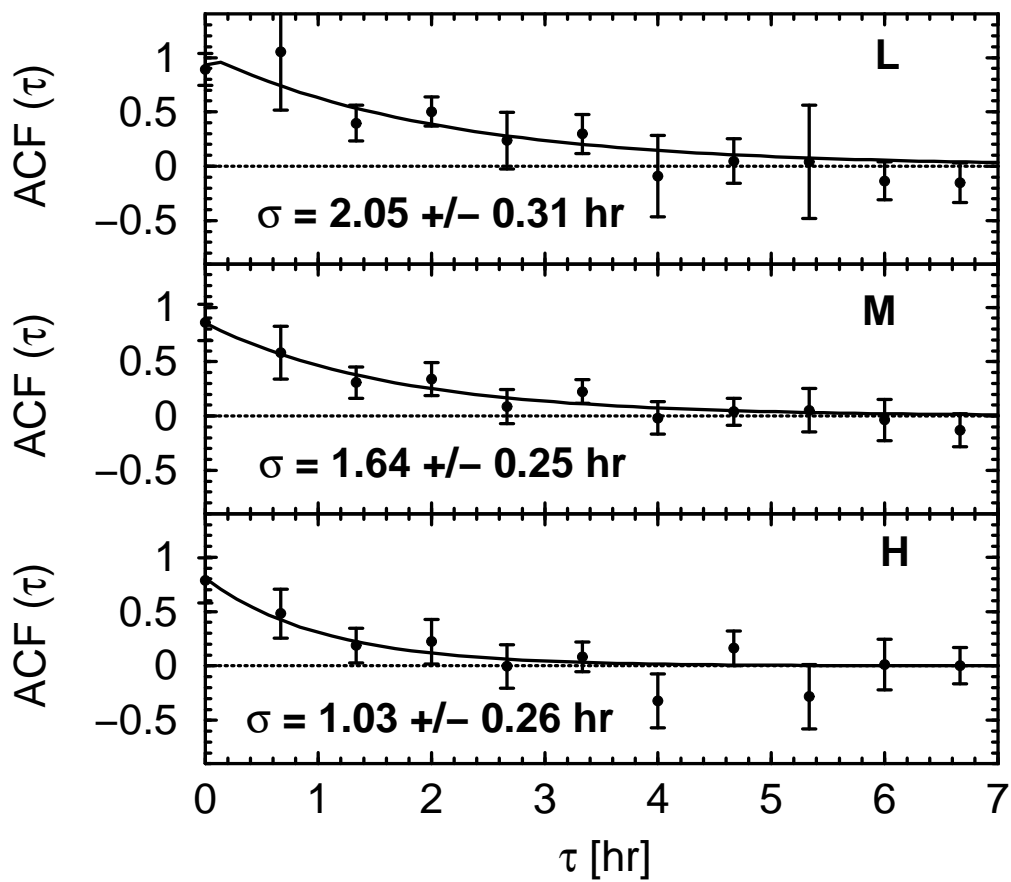}
\caption{Discrete autocorrelation functions (ACFs) of the X-ray flux in three 
energy channels: L = LECS (0.5 - 2)~keV, M = MECS (2 - 4)~keV, H = MECS (4 - 10)~keV. 
The ACFs have been fitted with a symmetric constant + exponential. $\sigma$ as 
quoted in the individual panels is the best-fit value of the decay time constant.}
\label{sax_acf}
\end{figure}

\cite{ravasio03} have defined a minimum doubling time scale
$T_{\rm short}$ to quantify the energy dependence of the short-term
variability time scale, and found no significant trend of this
quantity with photon energy. For all three energy LECS + MECS 
energy channels the minimum doubling time scales were found to 
be consistent with values of $\sim 6$~ksec. An estimate of the 
average rise and decay time scales in the rapid variability can 
be found through the width of the autocorrelation function (ACF) 
of the light curves. We have calculated the discrete autocorrelation 
functions for the three LECS + MECS light curves, and fitted them 
with an exponential. The results are plotted in Fig. \ref{sax_acf} 
and suggest a decreasing trend of the variability time scale with 
increasing photon energy, as illustrated in Fig. \ref{sax_acfwidths}. 
Clearly, the statistical errors on these measurements are too large
to seriously constrain the functional dependence of the ACF widths 
on photon energy. However, in \S \ref{parameter_estimates} we suggest
a new method to use a decreasing ACF width with increasing photon energy
for an independent magnetic-field estimate, and apply this method to
the {\it Beppo}SAX results, tentatively taking the best-fit results
at face value.

\subsection{\label{gammarays}Gamma-ray observations}

BL~Lacertae has been observed by the HEGRA system of imaging
Cherenkov telescopes, accumulating a total of 10.5~h of on-source
time in Sept. -- Nov. 2000. The source was not detected above a
99~\% confidence-level upper limit of 25~\% of the Crab flux at photon 
energies above 0.7~TeV \citep{mang01}. Assuming an underlying
power-law with energy spectral index $\alpha$, this corresponds
to a $\nu F_{\nu}$ flux limit of $8.65 \times 10^{11} \, \alpha$~Jy~Hz
at 0.7~TeV. 

\section{\label{variability}Spectral variability}

In this section, we will describe local spectral variability
phenomena, i.e. the variability of local spectral (and color) 
indices and their correlations with monochromatic source fluxes. 

\subsection{\label{B_minus_R}Optical spectral variability}

The optical spectral variability of BL~Lacertae during our campaign
has been investigated in great detail by \cite{villata02}. In the
following, we briefly summarize their results. \cite{villata02} have
calculated the de-reddened, host-galaxy subtracted B - R color indices
for a set of 620 observations taken by the same instrument within 20
minutes of each other, and with individual errors of the B and R
magnitudes of no more than 0.04 and 0.03 mag, respectively. Obvious
spectral variability was detected, and the color changes were more
sensitive to rapid variations than the long-term flux level. During
well-sampled, short flares (on time scales of a few hours), the 
color changes strictly follow the flux variability in the sense 
that the spectra are harder when the flux is higher \citep[see Fig. 
7 of][]{villata02}. A plot of B - R vs. R reveals two separate 
regimes within which the R magnitudes are well correlated with 
the respective B - R colors. However, there seems to be a 
discontinuity at $R \sim 14$~mag, separating a high-flux and 
a low-flux regime. Within each regime, a similar range of B - R 
colors is observed. \cite{villata02} have subsequently fitted 
the overall long-term flux variability by a cubic spline to 
the 10-day averages of the R-band light curve and rescaled 
this spline to pass through the minima of the light curve.
Cleaning the B and R fluxes from this base-flux level, they removed 
the long-term variability from the color-intensity correlation,
and found a very clean correlation between the superposed short-term
R-band variability and the B - R spectral hardness. This strongly
suggests that the optical long-term flux variability (on time scales
of weeks) is due to an achromatic mechanism, while the rapid (intraday)
variability is clearly chromatic \citep{villata02}.

\subsection{\label{X-ray_spectral_var}X-ray spectral variability}

Fig. \ref{rxte_flux_alpha_lc} displays the history of the best-fit
spectral index from the {\it RXTE} PCA monitoring observations, along
with the PCA light curve. Due to the relatively short exposure times,
the errors on the spectral indices are rather large, but a general
trend of the local spectral index being softer during strong hard X-ray 
(2 -- 10~keV) flares is discernible. This applies to the overall
low state (see, e.g., the flares at MJD 51708 = June 13 or MJD~51770 
= Aug.~14) as well as to the high state (e.g., MHD~51826 = Oct.~9 or
MJD~51853 = Nov.~5). In order to investigate the question of a 
hardness-intensity correlation on the timescale of a few days probed 
by the {\it RXTE} monitoring, we have constructed a hardness-intensity
diagram from the PCA data. To assess the average properties of 
the source at low fluxes, we have rebinned the points included 
in Fig.\ref{rxte_flux_alpha_lc} in flux bins of $\Delta F_{2 - 10} 
= 10^{-12}$~ergs~cm$^{-2}$~s$^{-1}$ for all individual points with 
2 -- 10~keV flux of less than $1.8 \times 10^{-11}$~ergs~cm$^{-2}$~s$^{-1}$,
calculating average fluxes and spectral indices weighted by the
inverse of the errors of the spectral-index measurements. Data 
points with larger 2 -- 10~keV fluxes are plotted individually. 
The result is shown in Fig. \ref{rxte_hid}. The figure 
illustrates that high PCA 2 -- 10~keV fluxes above $\sim 2 \times 
10^{-11}$~ergs~cm$^{-2}$~s$^{-1}$ are always characterized by soft
spectra with $\alpha \ge 1$, indicating decaying $\nu F_{\nu}$ spectra. 
This confirms the notion mentioned earlier that large-amplitude X-ray 
flares might be dominated by the variability of the low-energy (synchrotron) 
component, extending into the PCA energy range during flaring 
activity. No significant trend of the local spectral index with 
X-ray flux is discernible at low X-ray flux states.

The X-ray spectral variability on short (intra-day) time scales can
be characterized through variations of the {\it BeppoSAX} hardness 
ratios HR1 and HR2 as defined in \S \ref{nov1}. Their history during
the second {\it BeppoSAX} observation on Oct. 31 -- Nov. 2 is plotted
along with the LECS and MECS light curves in Fig. \ref{sax_lc}. Considering
the entire Oct. 31 -- Nov. 2 observation, we only find a very weak
hint of an anti-correlation of HR1 with the soft LECS (0.5 -- 2)~keV 
flux and a positive correlation of HR2 with the medium-energy MECS 
(4 -- 10)~keV flux. These trends become slightly more apparent when
following individual, well-resolved flares. Figs. \ref{loops_45}
and \ref{loops_48} show two examples of such hardness-intensity
diagrams for the flares around 45~hr and 48~hr of Oct. 31 (see Fig.
\ref{sax_lc}). A weak hardness-intensity anti-correlation at soft
X-rays (HR1 vs. LECS) and a positive hardness-intensity correlation 
at medium-energy X-rays (HR2 vs. MECS) can be seen. The flare around 
48~hr also shows weak evidence for spectral hysteresis as found 
previously in several HBLs such as Mrk 421 and PKS~2155-304 
\citep[e.g.,][]{takahashi96, kataoka00}. Such spectral hysteresis 
phenomena have been modelled in detail with pure SSC models for the
case of HBLs \citep[e.g.,][]{krm98,gm98,kataoka00,ktl00,lk00} and 
recently also predicted for intermediate and low-frequency peaked 
BL~Lacs by \cite{bc02}. 

\section{\label{crosscorrelations}Inter-band cross-correlations and time lags}

In this section, we investigate cross-correlations between the
measured light curves at different frequencies, within individual
frequency bands as well as broadband correlations between different
frequency bands. 

\subsection{\label{radio_crosscorrelations}Radio correlations}

We have calculated the discrete correlation functions (DCFs; \cite{ec88})
between the light curves at different radio frequencies for a variety
of sampling time steps $\Delta \tau$. We did not find any conclusive 
hints of correlations of the 37~GHz light curve with other radio 
light curves or light curves at other wavelength bands. This might
be the result of the poor sampling of the 37~GHz light curve.
Fig. \ref{dcf_radio} displays the DCFs of the various radio light 
curves at $\nu < 22$~GHz with the 14.5~GHz light curve as reference
for a sampling time scale of $\Delta\tau = 5$~d. We chose the 14.5~GHz 
reference light curve because it is the best sampled radio light curve 
in our data set. The DCFs show evidence for a correlation between 
the variability at the various radio frequencies. We have subsequently 
fitted the DCFs with Gaussians to determine the most likely time delays 
between the signals at different radio frequencies. We have repeated 
this procedure for several other sampling time scales (specifically, 
$\Delta\tau = 3$~d and $\Delta\tau = 7$~d) and found that only the 
result pertaining to the 14.5~GHz vs. 22~GHz DCF remained robust,
indicating a time lag between the 14.5~GHz and the 22~GHz light 
curve, with the 14.5~GHz light curve lagging behind the 22~GHz one 
by $\sim 15$~d. The best fit is indicated by the solid curve in the 
top panel of the figure. 

In order to test the statistical significance of the high-frequency
radio time lag, we have performed a series of 10,000 Monte-Carlo
simulations, assuming uncorrelated variability patterns between
the 14.5~GHz and the 22~GHz light curves. Specifically, we have
simulated random light curves for the 22~GHz fluxes and performed
the same DCF and Gaussian fitting analysis as we had done with the
actual data. The simulated 22~GHz light curves were constrained
by the measured maximum and minimum fluxes in our data set and by
a short-term doubling time scale of 40~d for rapid fluctuations on 
$\Delta t \le 10$~d, and a long-term doubling time scale of 4~mo.
Simulated data points were constructed for the times of the actual 
22~GHz measurements in our data set. We find a probability of 
12.3~\% that these simulated random light curves show a DCF 
amplitude higher than resulting from the real 22~GHz flux history, 
with an acceptable Gaussian fit to the DCF. Thus, the measured 
time delay can not be considered statistically significant. This 
might be, at least in part, due to the fact that the data train 
from our campaign is relatively short.

\subsection{\label{radio_optical}Radio -- optical correlations}

We also calculated the DCFs between the light curves at the various 
radio and the optical (R-band) light curves, and calculated the best-fit
time delay, as described in the previous paragraph. This resulted in 
low-significance detections of time delays of $\sim 45$ -- 50~d of the 
8, 14.5, and 37~GHz radio light curves behind the optical ones. However, 
these results have to be considered with great caution since our data 
set only spans about half a year. On the basis of historical data, 
characteristic time delays of $\sim 1$ -- 4.5 years between the 
radio and optical variability had been found previously \citep{bregman90}. 
Thus, our correlations may well be a chance coincidence in the sense 
that the observed radio and optical variability patterns may not 
correspond to the same epoch of activity of the central source. A 
more comprehensive study of the long-term behaviour of BL~Lacertae, 
including the results of this multiwavelength campaign will be the 
subject of subsequent work (e.g., \cite{villata03}). 

As reported by \cite{villata02}, the light curves in the different optical
bands (U, B, V, R, I) are well correlated (with the hardness - brightness
correlation described in \S \ref{B_minus_R}), but no significant, measurable
time delays between the B and the R band (the best sampled optical light
curves in our campaign) were found.

\subsection{\label{x_optical}X-ray -- optical correlations}

The extraordinary time coverage of the R band light curves during the
core campaign allows us to do a meaningful comparison of the intraday
variability patterns at optical and X-ray frequencies during our first
{\it Beppo}SAX observation of July 26/27. Fig. \ref{r_lecs_lc} displays
the LECS [0.7 -- 2]~keV and the contemporaneous R-band light curve
during this observation. From visual inspection, it seems that the 
R-band light curve closely tracks the LECS light curve with 
a delay of $\sim 4$ -- 5~hr (indicated by the dotted arrows in Fig.
\ref{r_lecs_lc}). However, the DCF between the R-band flux and the 
LECS count rate (see Fig. \ref{r_lecs_dcf}) instead identifies a 
stronger apparent signal from an anti-correlation with an optical -- 
X-ray delay of $\sim 3$~hr. The dominant features probably identified 
by the DCF analysis are indicated by the dot-dashed arrows in Fig.
\ref{r_lecs_lc}. Note that the DCF is defined so that negative $\tau$ 
corresponds to a lead of the X-rays. Unfortunately, the limited statistics
of the {\it Beppo}SAX light curve prevents a more in-depth analysis of
the possible optical -- X-ray correlation on these short time scales.
Any claimed correlation does not hold up to a statistical significance
test. However, if we assume that the optical lag of $\sim 4$ -- 5~hr is
real and can be interpreted as due to synchrotron cooling, it allows 
an independent magnetic field estimate, which will be quantified 
in \S \ref{parameter_estimates}. Interestingly, the resulting magnetic
field is in good agreement with an independent estimate from a basic 
equipartition argument.

The DCF also identifies a strong, apparent correlation between optical 
and X-ray fluxes with an R-band lead of $\sim 9$~hr, which seems to
arise from the large optical flare at $\sim 8$~hr, preceding the LECS 
flare at $\sim 17$~hr (see long-dashed arrows in Fig. \ref{r_lecs_lc}). 
However, we believe that this might be an artifact due to the limited 
duration and time resolution of the LECS light curve. Note that this 
9~hr time scale spans over about half the duration of the entire
{\it BeppoSAX} LECS light curve for this observation. We have also looked 
for correlations between the R band and X-ray fluxes on longer time scales, 
applying the DCF analysis to the {\it RXTE} PCA, ASM (daily averages), 
and R-band light curves. No significant time delays were detected.

We have also investigated possible time delays between the different
{\it Beppo}SAX LECS and MECS light curves displayed in Fig. \ref{sax_lc}
for the Oct. 31 -- Nov. 2 observation. While the DCFs show evidence for 
a positive correlation between all of these light curves, no measurable 
time delays could be identified.

\section{\label{spectra}Broad-band spectral energy distributions}

We have constructed simultaneous broadband spectral energy distributions
(SEDs) for the times of the two {\it Beppo}SAX pointings. They are shown
in Fig. \ref{mw_combined}. For the radio fluxes, we selected the measurements 
closest in time to the center of the respective {\it Beppo}SAX exposure. We 
generally had multiple optical flux measurements throughout the times of the 
{\it Beppo}SAX exposures. To construct the SEDs, we have calculated the average 
optical flux in each band from the de-reddened, host-galaxy-subtracted 
individual flux measurements (see \S \ref{optical}) over the {\it Beppo}SAX 
exposure time, and indicate the (in some cases rather substantial) range of 
variability over that period by the error bars on the optical fluxes. 

We represent the best-fit power-law spectra of the {\it Beppo}SAX
LECS + MECS measurements as well as the simultaneous or quasi-simultaneous
{\it RXTE} PCA measurements as bow-tie outlines in the SEDs. For
orientation purposes only, we also indicate the EGRET flux from
the major $\gamma$-ray outburst of BL~Lacertae in July 1997
\citep{bloom97}. Included are also the anticipated sensitivity
limits of current and future atmospheric \v Cerenkov telescope
facilities, and the HEGRA upper limit from the observations in
Sept. -- Nov., 2000.

Fig. \ref{mw_combined} illustrates the drastically different activity
states between the July 26/27 and the Oct. 31 -- Nov. 2 {\it Beppo}SAX
observations. In the July 26/27 SED, the synchrotron peak appears
to be located at frequencies clearly below the optical range (probably
at $\nu_{\rm sy} \sim 10^{14}$~Hz), and the synchrotron emission cuts
off at a frequency near or below $\sim 10^{17}$~Hz. In contrast, the
SED of Oct. 31 -- Nov. 2 shows clear evidence for the presence of the
synchrotron component out to at least 10~keV, and the synchrotron peak
might be located in the optical range at a few times $10^{14}$~Hz. For
illustration purposes, we have also plotted the {\it RXTE}~PCA spectrum
of the observation a few hours before the beginning of the Oct. 31 --
Nov. 2 {\it Beppo}SAX pointing. This PCA spectrum shows characteristics
rather similar to the low-state spectrum, and illustrates the drastic
nature of the short-term X-ray variability. This might indicate that
the rapid variability of BL~Lacertae is probably driven by short episodes
of injection / acceleration of high-energy electrons into the emitting
volume. 

\cite{ravasio03} have shown that the extrapolation of the optical spectrum 
towards higher frequencies does not connect smoothly with the contemporaneous 
soft X-ray spectrum (see their Fig. 5). They have considered various
possible explanations of this discrepancy: (a) a variable dust-to-gas
ratio, which could cause a larger degree of reddening during the 
Oct. 31 -- Nov. 2 observations, (b) the Bulk-Compton process \citep{sikora97},
which could provide an additional emission component at soft X-rays due
to Compton upscattering of external photons by a thermal component of
electrons in the emitting volume, (c) a second component of relativistic
electrons, providing an additional source of synchrotron emission at
soft X-rays, and (d) the flattening effect of the Klein-Nishina cutoff
on the cooling electron distribution, leading to a high-energy bump
in the synchrotron emission \citep[see][for a recent application of
this idea to the optical -- X-ray spectra of individual knots in
jets of radio galaxies detected by {\it Chandra}]{da02}. While the 
first two of the ideas listed above were found to lead to rather 
unrealistic inferences about the environment of the AGN and the 
underlying accretion-disk spectrum, respectively, the latter two 
scenarios will be considered further in our modelling efforts in 
our companion theory paper (B\"ottcher \& Reimer 2003, in preparation). 
However, it seems also possible that this misalignment could be an 
artifact of the flux averaging over the $\sim 1.5$~days of the 
{\it Beppo}SAX observations, including multiple short-term flares 
of only a few hours each. In order to test for this possibility, 
it will be essential to use a fully time-dependent AGN emission 
model and do the flux averaging in a similar way as was done with 
the data to construct the SEDs displayed in Fig. \ref{mw_combined}.

\section{\label{parameter_estimates}Generic parameter estimates}

In this section, we discuss some general constraints on source 
parameters that will be relevant for detailed spectral and variability 
modelling of BL~Lacertae. We will first (\S \ref{magnetic_field}) 
focus on three independent methods to estimate the magnetic field, 
and then turn to other properties, such as the co-moving Lorentz 
factors of electrons in the jet, the bulk Lorentz and Doppler 
boosting factors, the kinetic luminosity of the jet, and the 
source size.

\subsection{\label{magnetic_field}The magnetic field}

\begin{figure}
\plotone{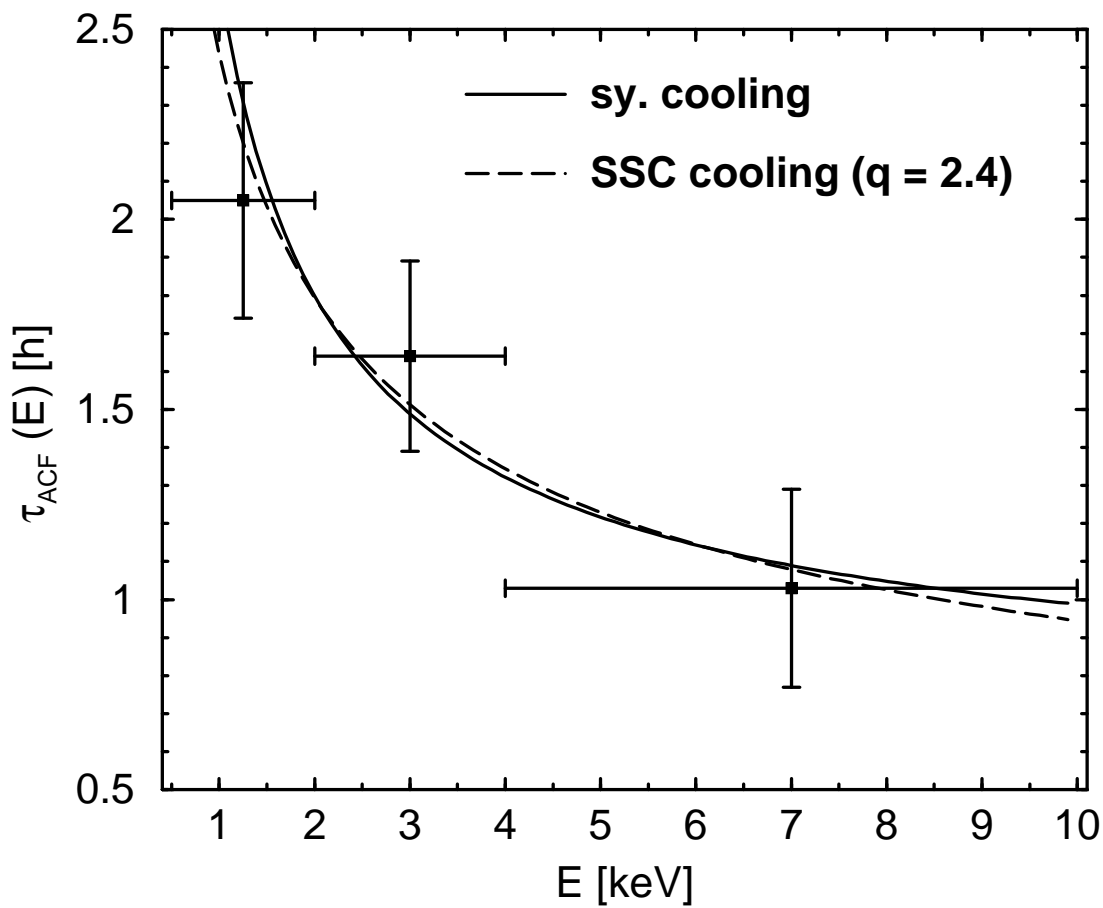}
\caption{The widths of the discrete autocorrelation functions (ACFs) of the light 
curves in the LECS + MECS energy range, as a function of synchrotron photon energy. 
The solid curve is the best fit of this energy dependence assuming it is caused
by synchrotron + external-Compton cooling; the solid curve is a fit assuming
dominant SSC cooling with an injection electron spectral index of $q = 2.4$.}
\label{sax_acfwidths}
\end{figure}

\begin{figure}
\plotone{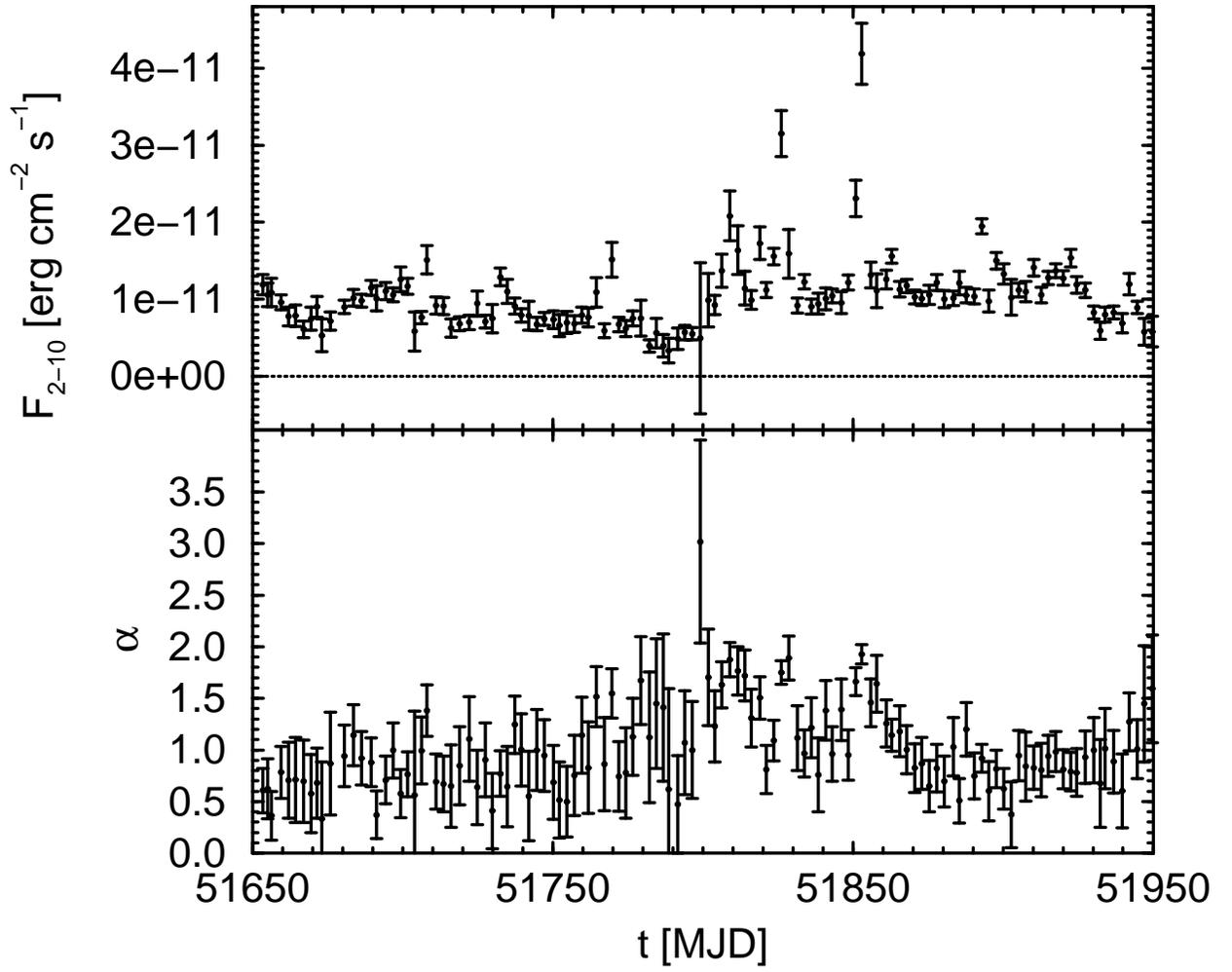}
\caption{RXTE PCA flux and best-fit spectral-index history.}
\label{rxte_flux_alpha_lc}
\end{figure}

\begin{figure}
\plotone{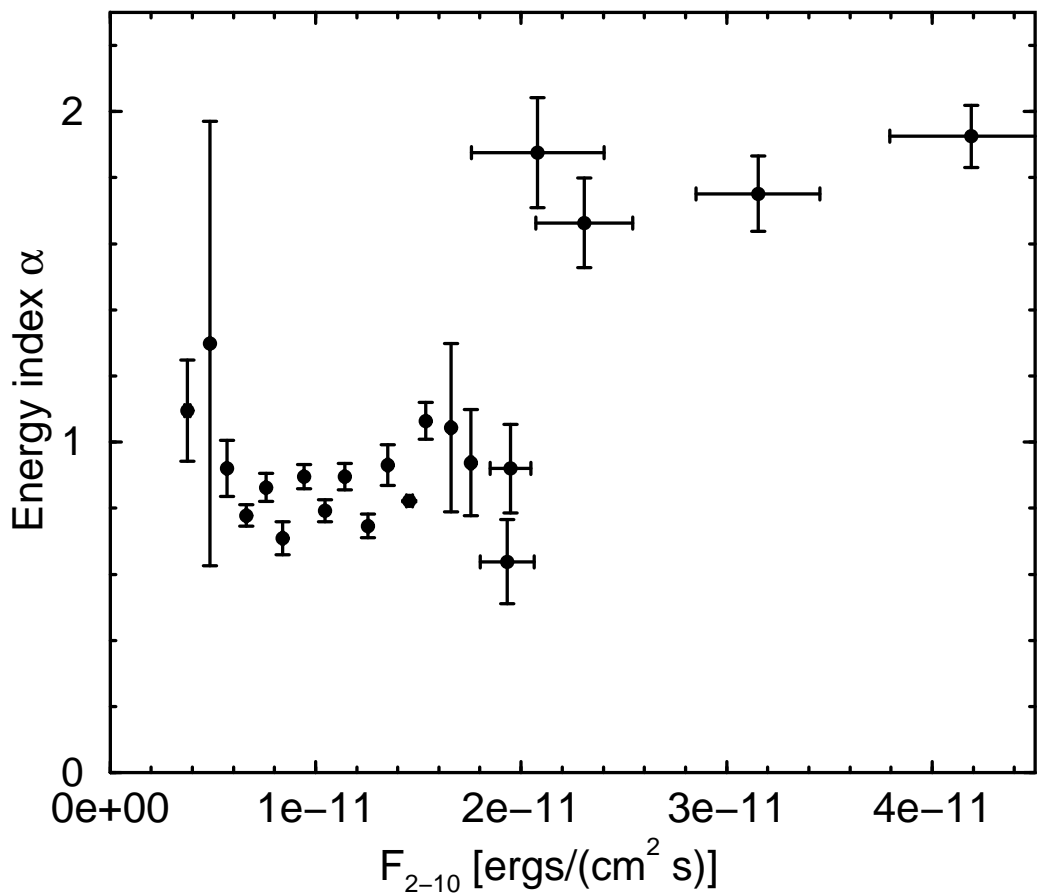}
\caption{Hardness-intensity plot for the {\it RXTE} PCA measurements. At $F_{2-10}
< 1.8 \times 10^{-11}$~ergs~cm$^{-2}$~s$^{-1}$, the data have been rebinned into
flux bins of $\Delta F_{2-10} = 10^{-12}$~ergs~cm$^{-2}$~s$^{-1}$. At higher flux
values, data points from individual PCA observations are displayed. High flux
states are always associated with a soft spectrum, but no obvious hardness-intensity
correlation is visible at low X-ray flux levels.}
\label{rxte_hid}
\end{figure}

\begin{figure}
\plotone{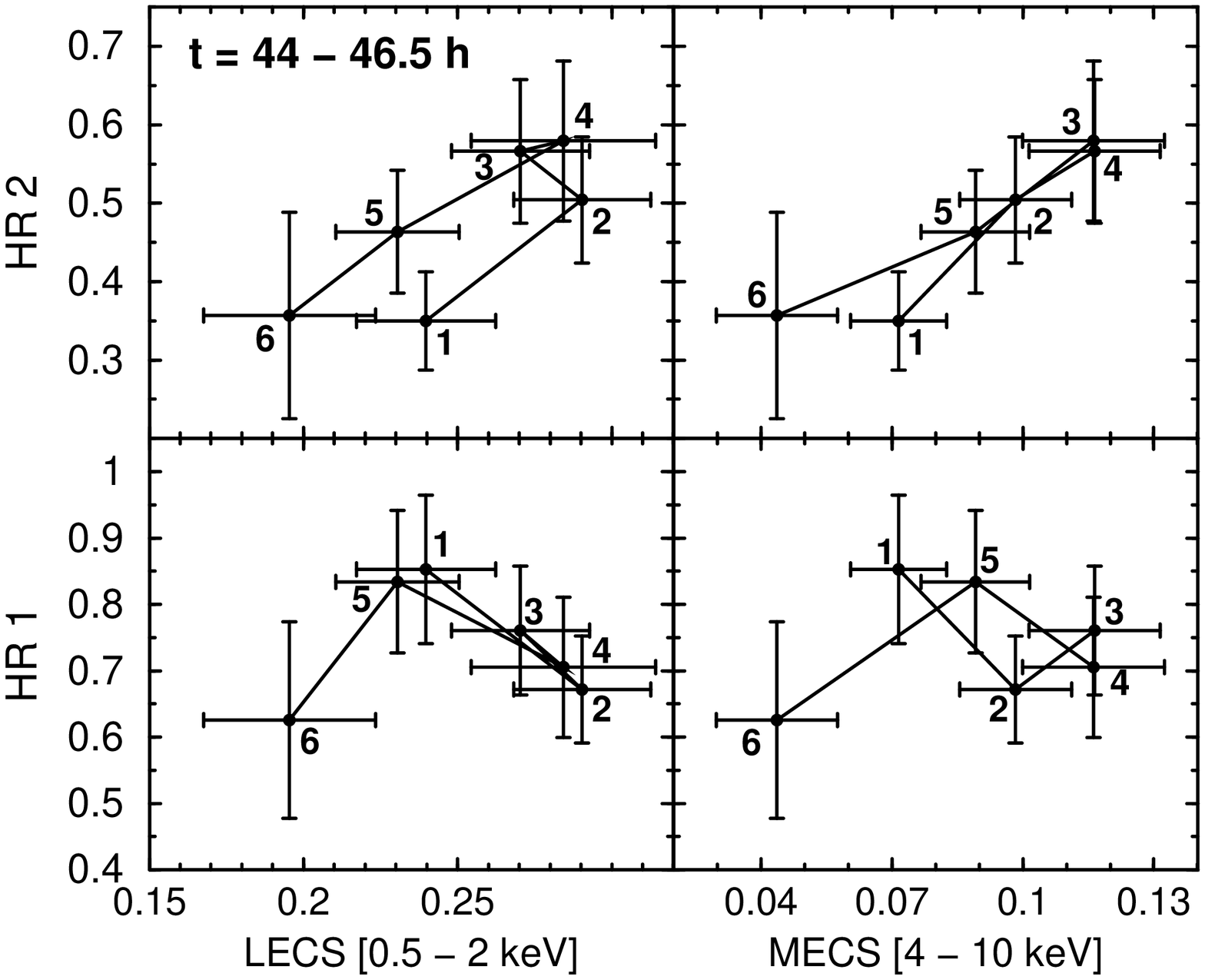}
\caption{Hardness-intensity diagram of the {\it BeppoSAX} hardness ratios
HR1 and HR2 as defined in \S \ref{nov1} vs. soft X-ray LECS and medium-energy
MECS flux for the well-resolved X-ray flare at t = 44.0 -- 46.5~h
(see Fig. \ref{sax_lc}).}
\label{loops_45}
\end{figure}

\begin{figure}
\plotone{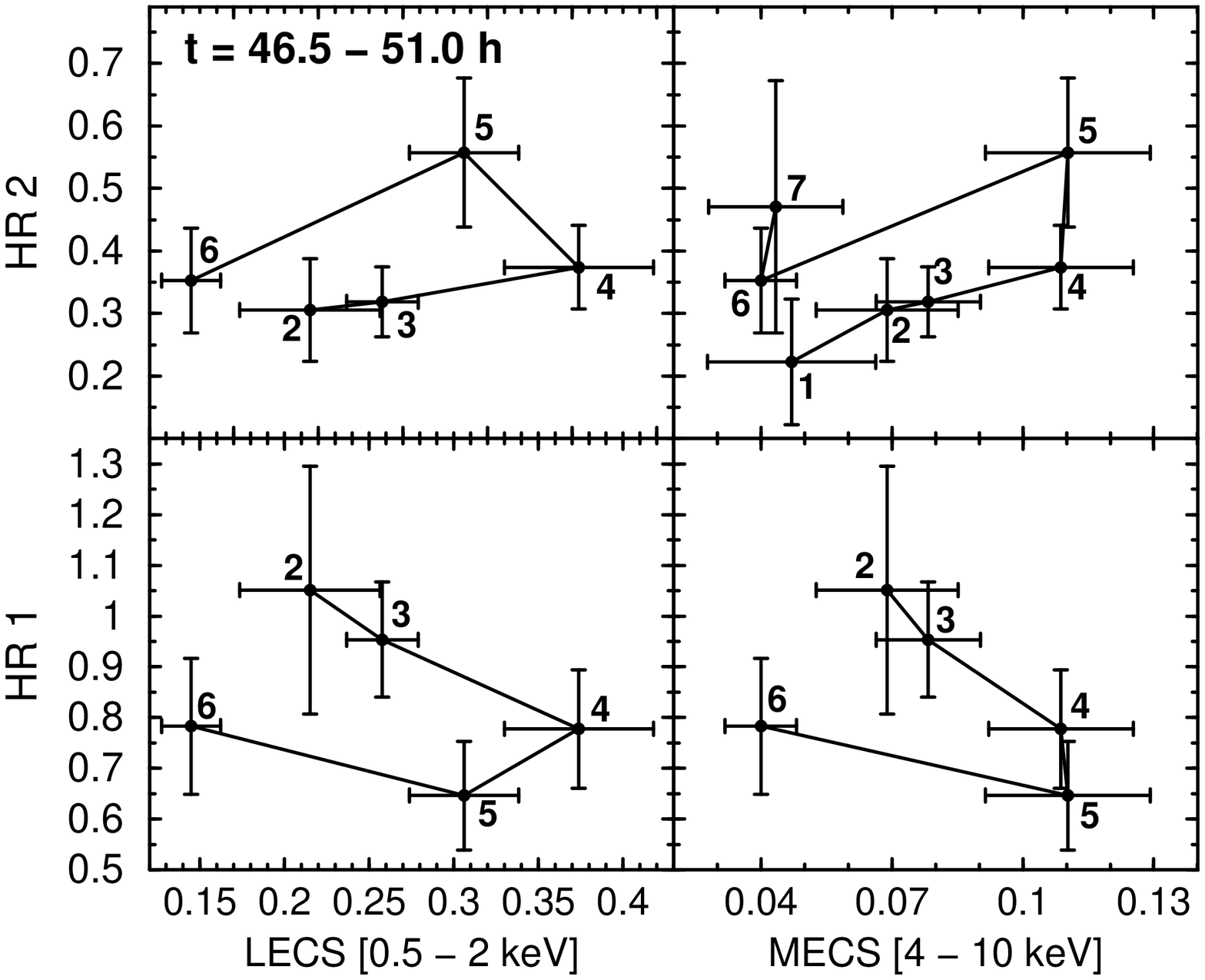}
\caption{Hardness-intensity diagram of the {\it BeppoSAX} hardness ratios
HR1 and HR2 as defined in \S \ref{nov1} vs. soft X-ray LECS and medium-energy
MECS flux for the well-resolved X-ray flare at t = 46.5 -- 51.0~h
(see Fig. \ref{sax_lc}).}
\label{loops_48}
\end{figure}

\begin{figure}
\plotone{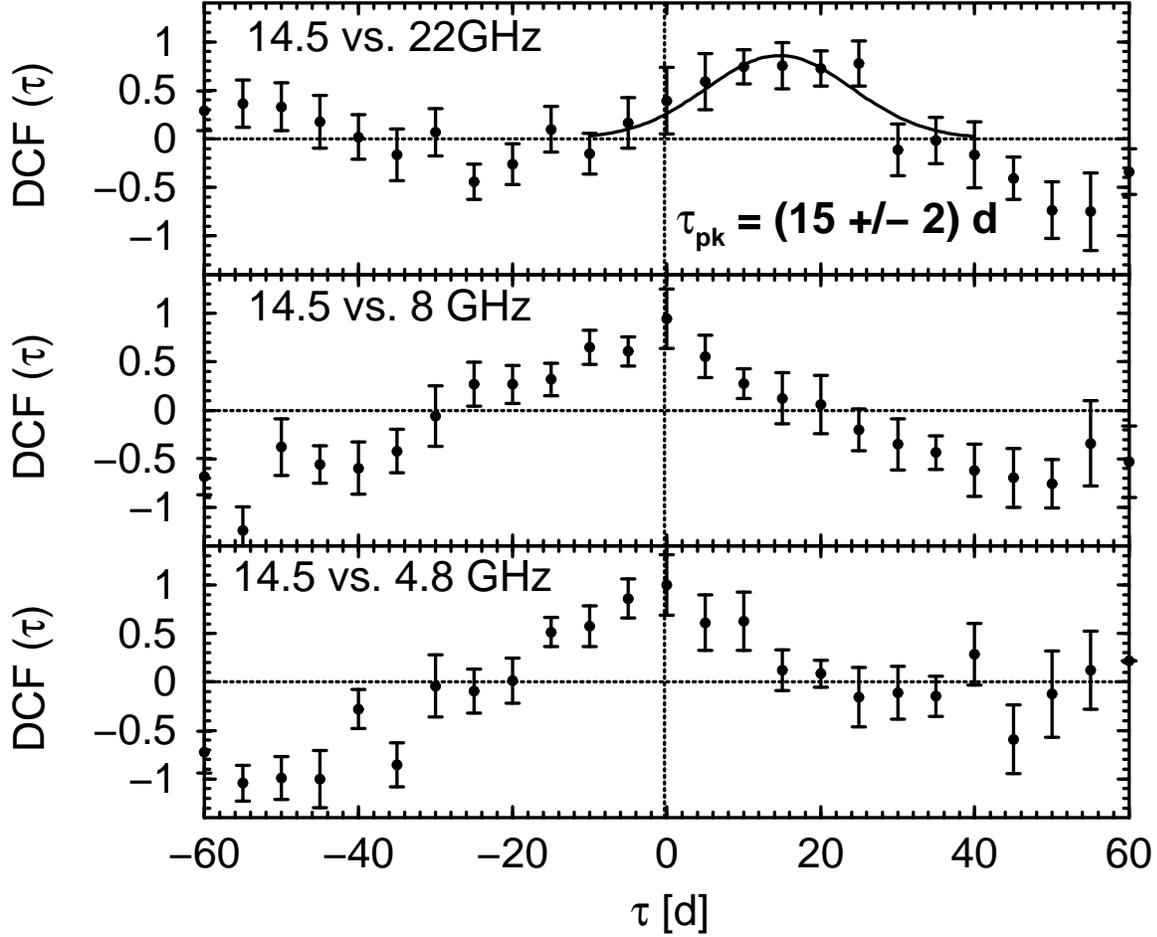}
\caption{Discrete correlation functions (DCFs) between the 14.5~GHz radio 
light curve (the best sampled radio light curve available in our data set) 
with respect to light curves at other radio frequencies. The solid curve 
in the top panel indicates a Gaussian fit to the DCF, with $\tau_{\rm pk}$ 
being the best-fit offset from zero (i.e., the most likely time delay). 
This is only included for the 14.5~GHz vs. 22~GHz DCF since this procedure 
did not produce robust results (independent of the sampling time scale) for 
the other DCFs shown.}
\label{dcf_radio}
\end{figure}

\begin{figure}
\plotone{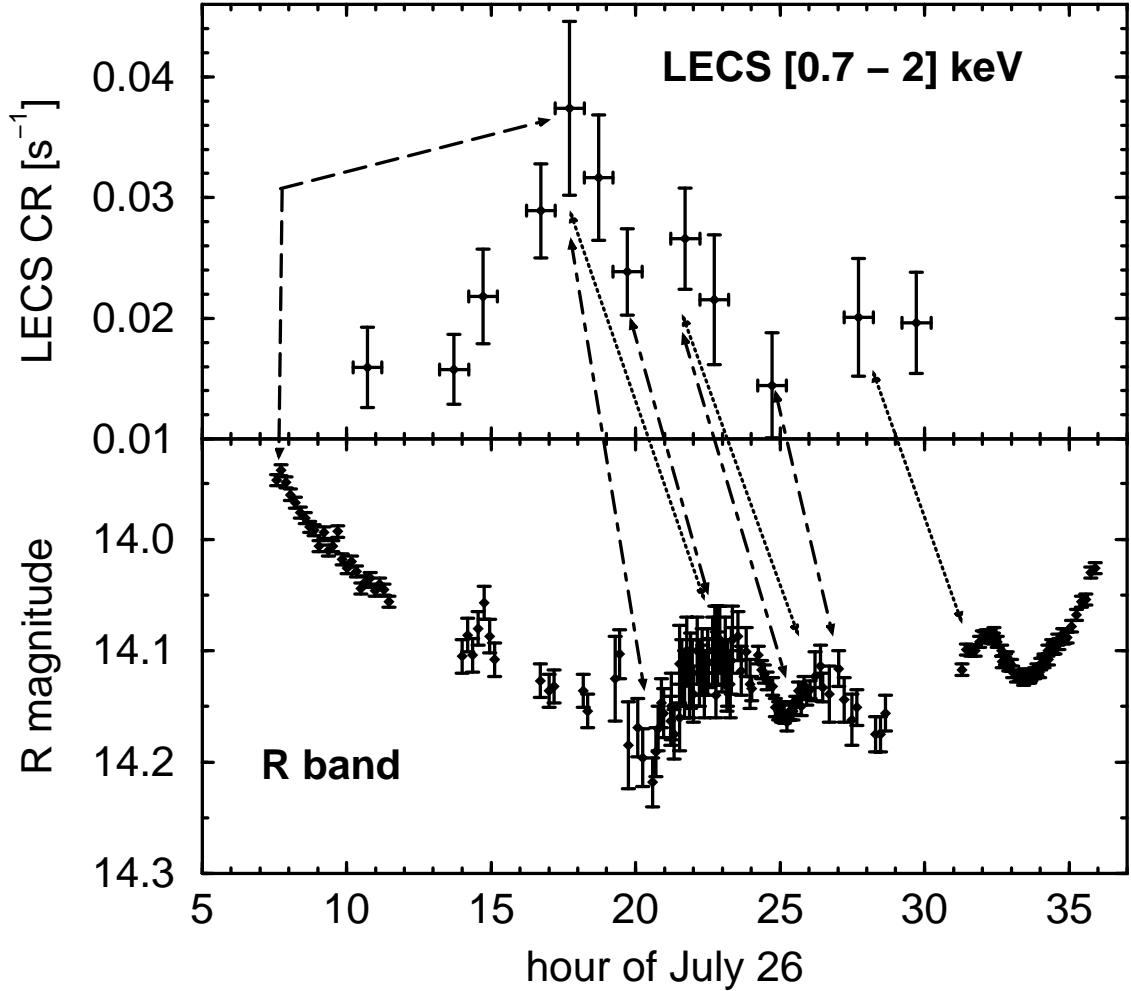}
\caption{{\it Beppo}SAX LECS [0.7 -- 2]~keV (top panel) and contemporaneous
R-band (bottom panel) light curves during our first {\it Beppo}SAX pointing
on July 26/27. The optical light curve appears to trace the X-ray light
curve with a time delay of $\sim 4$ -- 5~hr, as indicated by the dotted
arrows. However, a DCF analysis (see Fig. \ref{r_lecs_dcf}) actually finds 
a stronger signal from an apparent anti-correlation with a time delay of 
$\sim 3$~hr, dominated by features indicated by the dot-dashed arrows as
well as a strong positive correlation with an X-ray lag of $\sim 9$~hr
behind the optical, dominated by the large optical and X-ray flares 
indicated by the long-dashed arrow.}
\label{r_lecs_lc}
\end{figure}

\begin{figure}
\plotone{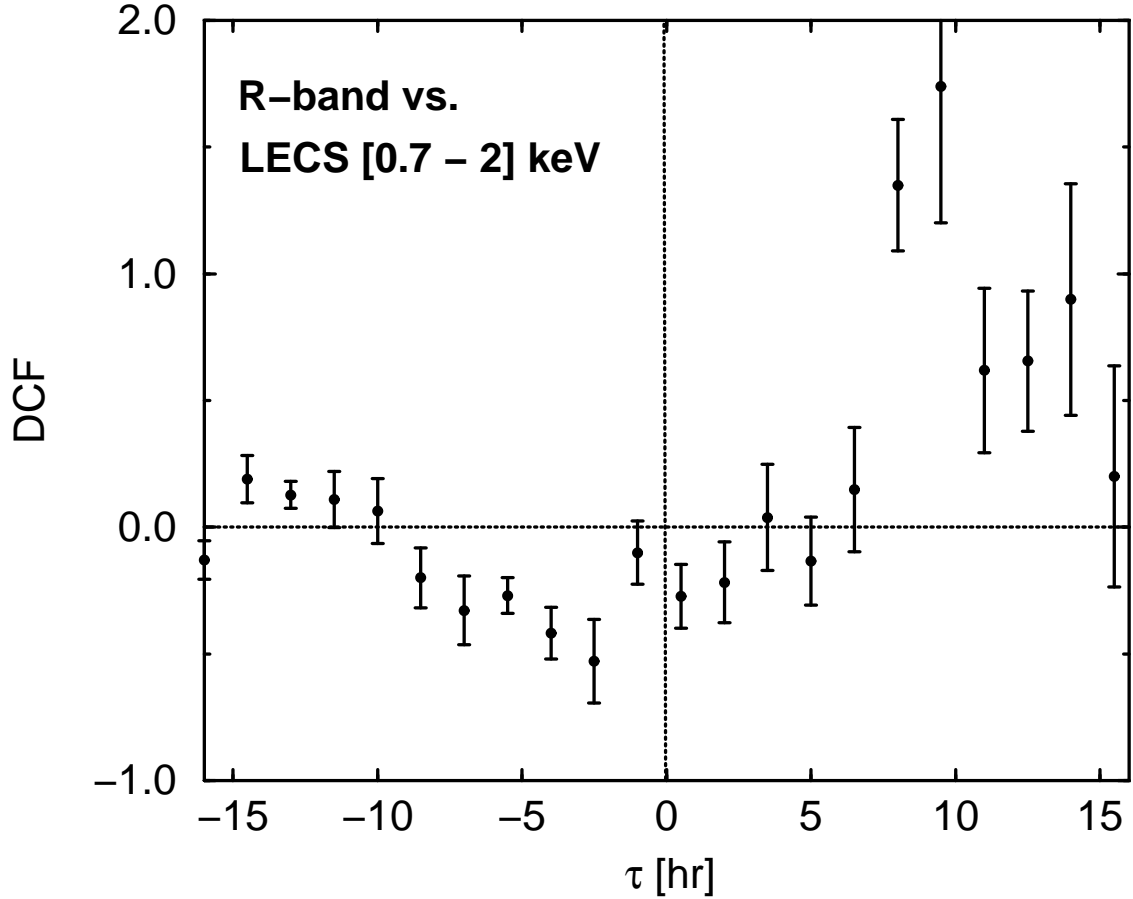}
\caption{Discrete correlation function between the {\it Beppo}SAX LECS 
[0.7 -- 2]~keV and R-band light curves during our first {\it Beppo}SAX 
pointing on July 26/27, as shown in Fig. \ref{r_lecs_lc}. The sampling
time scale is $\Delta\tau = 1.5$~hr. The same general features are found
for other sampling time scales in the range of $\sim 1$ -- 2~hr. }
\label{r_lecs_dcf}
\end{figure}

\begin{figure}
\plotone{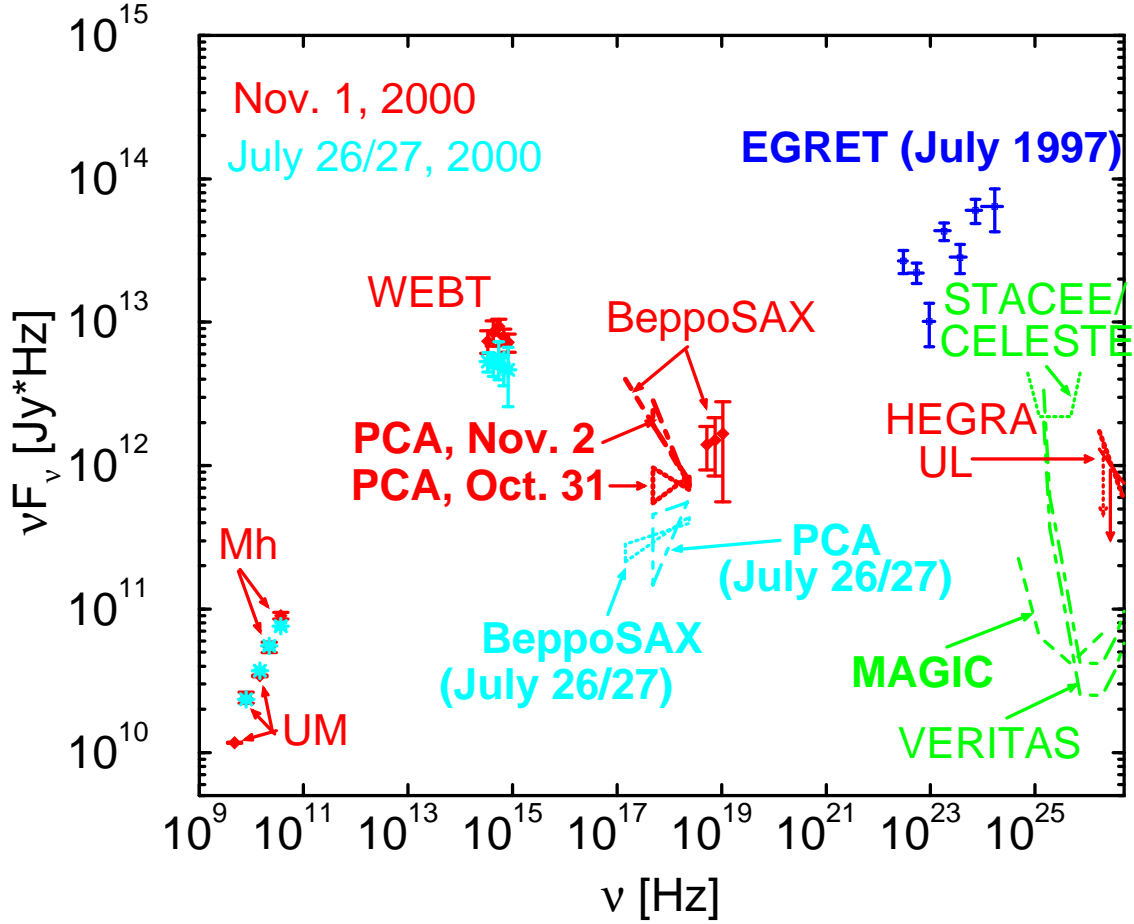}
\caption{Spectral energy distributions of BL~Lacertae on July 26/27, 2000
(stars; cyan in the on-line version; light grey in print), and Oct. 31 -- 
Nov. 2, 2000 (diamonds; red in the on-line version; dark grey in print). 
For Oct. 31 -- Nov. 2, both the PCA measurement a few hours before the 
{\it Beppo}SAX pointing, and near the end of the {\it Beppo}SAX pointing 
are included. ``Mh'' indicates the radio fluxes measured at Mets\"ahovi 
Radio Observatory; ``UM'' indicates the University of Michigan Radio 
Observatory measurements. The two VERITAS sensitivity limits have been 
obtained assuming two different values of the underlying spectral index, 
$\alpha = 2.5$, and $\alpha = 3.5$, respectively.}
\label{mw_combined}
\end{figure}

In \S \ref{xrays}, we had investigated the energy-dependent 
width of the autocorrelation functions of the X-ray variability
during the Oct. 31 -- Nov. 2 {\it Beppo}SAX observation (see
Fig. \ref{sax_acfwidths}). Assuming that the rise time of short-term 
flaring is not significantly dependent on energy (i.e., it is 
dominated by light-crossing time constraints rather 
than an energy-dependent acceleration time scale), the 
width of the ACF should yield an estimate of the cooling 
time scale $\tau_c (E)$ of the electrons responsible for the
emission at energy $E$ as a function of photon energy. 
Specifically, in that case, we expect that $\tau_{\rm ACF} 
(E) = \tau_0 + \tau_{\rm cool} (E)$. If the LECS + MECS spectrum 
is indeed dominated by synchrotron emission, we can associate an 
observed photon energy $E = 1 \, E_{\rm keV}$~keV with the 
characteristic photon energy of synchrotron emitting 
electrons, $E_{\rm keV} = 1.4 \times 10^{-11} \, D \, B_{\rm G}
\, \gamma^2$, where $B = 1 \, B_{\rm G}$~G is the co-moving
magnetic field, and $\gamma$ is the electron Lorentz factor.
If the electron cooling is dominated by synchrotron and/or 
external Compton cooling in the Thomson regime, we have 
$\dot\gamma = - (4/3) \, c \, \sigma_T \, (u_B / m_e 
c^2) \, (1 + k) \, \gamma^2$, where $u_B$ is the energy 
density in the magnetic field, and $k \equiv u_{\rm ext} 
/ u_B$ is the ratio of the energy density in an external 
photon field to the magnetic-field energy density (all 
quantities in the co-moving frame of the emitting region). 
This yields the synchrotron + EC cooling time (in the 
observer's frame) of

\begin{equation}
\tau_{\rm cool, sy} (E) = 2.9 \times 10^3 \, D^{-1/2} 
B_{\rm G}^{-3/2} \, (1 + k)^{-1} \, E_{\rm keV}^{-1/2} 
\; {\rm s}.
\label{tau_sy}
\end{equation}
Fitting a function $\tau_{\rm ACF} (E) = \tau_0 + \tau_1 \,
E_{\rm keV}^{-1/2}$ to the energy dependence shown in Fig.
\ref{sax_acfwidths} (solid line) yields a best-fit value 
of $\tau_1 = (7800 \pm 1400)$~s, which implies $B = (0.24 
\pm 0.03) \, D_1^{-1/3} \, (1 + k)^{-2/3}$~G, where $D_1
= D/10$. 

Alternatively, if electron cooling is dominated by the
synchrotron-self-Compton process, the relevant photon
field energy density is $u_{\rm sy}$, which we can approximate
as

\begin{equation}
u_{\rm sy} \approx \tau_{\rm T} \, u_B \, {q - 1 \over 3 - q}
\, \gamma_1^{q - 1} \, \gamma_2^{3 - q},
\label{u_sy}
\end{equation}
where $\tau_{\rm T} = n_e \, \sigma_T \, R_{\rm B}$ is the
radial Thomson depth of the emitting region, and $q$ is the
spectral index of the {\it injected} electron spectrum, $Q(\gamma)
= Q_0 \, \gamma^{-q}$ for $\gamma_1 \le \gamma \le \gamma_2$. The 
decay time scale of the light curve at a characteristic synchrotron 
photon energy $E$ will then correspond to the Compton cooling 
time scale of electrons at the high-energy end of the electron 
spectrum, $\gamma_2$, at the time when their characteristic 
synchrotron frequency equals $E$. Thus, the relevant cooling 
rate is

\begin{equation}
{d\gamma_2 \over dt} = - {4 \over 3} c \, \sigma_T \, {U_B \over
m_e c^2} \, {q - 1 \over 3 - q} \, \tau_{\rm T} \, \gamma_1^{q - 1}
\, \gamma_2^{5 - q}.
\label{dg2dt}
\end{equation}
This yields a characteristic cooling time (in the observer's frame)
of

\begin{equation}
\tau_{\rm cool, SSC} = 7.7 \times 10^8 \, {3 - q \over q - 1} \,
(2.4 \times 10^5)^{q - 4} \, \tau_{\rm T}^{-1} \, \gamma_1^{1 - q}
\, B_{\rm G}^{- q/2} \, D^{(2 - q)/2} \; E_{\rm keV}^{(q - 4)/2}
\; {\rm s}.
\label{tau_ssc}
\end{equation}
Leaving the index $q$ free, we find a best-fit energy dependence
of $\tau_{\rm cool} = \tau_0 + \tau_1 \, E_{\rm keV}^{-\zeta}$ 
with $\zeta = 0.50 \pm 0.14$, which yields the same energy 
dependence as in the synchrotron-cooling case. However, the 
optical spectral index of $\alpha_o \sim 1.2$ corresponds to 
a time average (or equilibrium in the case of a balance between 
particle injection and radiative cooling) electron spectral 
index of $p = 3.4$, indicating a value of $q = 2.4$. Fixing 
$q = 2.4$ (i.e., $\zeta = 0.8$), the fit is still perfectly 
consistent with the measured energy dependence, as indicated 
by the dashed line in Fig. \ref{sax_acfwidths}. From the 
best-fit value of $\tau_1 = (7020 \pm 1980)$~s, we can derive
a magnetic-field estimate of $B = (0.20 \pm 0.05) \, \tau_{\rm T, -6}^{-5/6}
\, (\gamma_1 / 100)^{-7/6} \, D_1^{-1/12}$~G. With the available
data, the two cooling scenarios can obviously not be distinguished,
and the estimates may be rather uncertain due to the possible effect
of energy-dependent acceleration times and energy dependent photon
propagation times through the blob. Thus, our magnetic field estimates
based on Eqs. \ref{tau_sy} and \ref{tau_ssc} are meant more as a
suggestion for the analysis of higher-quality data from future observations
by X-ray observatories with higher throughput, like {\it Chandra} or 
{\it XMM-Newton} or the planned Constellation-X mission, rather than 
a realistic magnetic field estimate from our currently available 
{\it BeppoSAX} data.

Another independent magnetic-field estimate could be obtained from 
the time delay between the {\it Beppo}SAX LECS [0.7 -- 2]~keV and 
the R-band light curves of $\Delta t^{\rm obs} \sim 4$ -- 5~hr, for 
which Fig. \ref{r_lecs_lc} shows tantalizing, though not statistically 
significant, support. Assuming that the correlation is real and the
delay is caused by synchrotron cooling of high-energy electrons with
characteristic observed synchrotron photon energy $E_{\rm sy, 0} =
E_0$~keV to lower energies with corresponding synchrotron energy
$E_{\rm sy, 1} = E_1$~keV, we find a magnetic-field estimate 
analogous to Eq. \ref{tau_sy}:

\begin{equation}
B_{\rm delay} = 0.4 \, D_1^{-1/3} \, (1 + k)^{-2/3} \, 
(\Delta t_h^{\rm obs})^{-2/3} \, (E_1^{-1/2} - E_0^{-1/2})^{2/3}
\; {\rm G}.
\label{B_delay}
\end{equation}
where $\Delta t_h^{\rm obs}$ is the observed time delay in hours.
Using $\Delta t_h^{\rm obs} = 5$, $E_0 = 1$ for the LECS, and 
$E_1 = 5.6 \times 10^{-4}$ for the R band, we find 

\begin{equation}
B_{\rm delay, RX} = 1.6 \, D_1^{-1/3} \, (1 + k)^{-2/3} \; {\rm G}. 
\label{B_RX}
\end{equation}
We need to point out that Eq. \ref{B_RX} may, in fact, slightly 
overestimate the actual magnetic field since at least the optical
synchrotron emitting electrons may also be affected by adiabatic
losses and escape. Depending on the details (geometry and mechanism)
of the jet collimation, those processes can act on time scales as
short as the dynamical time scale, which is constrained by the 
observed minimum variability time scale of $\Delta t_{\rm dyn} 
\lesssim 1.5$~hr (in the observer's frame). Another note of caution
that needs to be kept in mind is that the rather large sampling 
time scale of the X-ray light curve of $\Delta t = 1$~hr, precludes
the estimation of magnetic fields larger than $B_{\rm delay, max}
\sim 4.8 \, D_1^{-1/3} \, (1 + k)^{-2/3}$~G from delays between 
the optical and X-ray light curves. Consequently, magnetic fields
larger than $B_{\rm delay, max}$ can not be excluded on the basis
of the present analysis.

In principle, one could also derive an analogous estimate for the 
SSC-cooling case. However, it is unlikely that SSC cooling can dominate 
when the bulk of the synchrotron emission has evolved down to optical 
frequencies. Thus, it would not be realistic to apply such an estimate 
to the optical -- X-ray time delay.

A third independent estimate of the co-moving magnetic field can
be found by assuming that the dominant portion of the time-averaged 
synchrotron spectrum is emitted by a quasi-equilibrium power-law 
spectrum of electrons with $N_e (\gamma) = n_0 \, V_B \, \gamma^{-p}$
for $\gamma_1 \le \gamma \le \gamma_2$; here, $V_B$ is the co-moving
blob volume. The normalization constant $n_0$ is related to the
magnetic field through an equipartition parameter $e_B \equiv u_B
/ u_e$ (in the co-moving frame). Note that this equipartition parameter
only refers to the energy density of the electrons, not accounting for
a (possibly greatly dominant) energy content of a hadronic matter 
component in the jet. Under these assumptions, the $\nu F_{\nu}$
peak synchrotron flux $f_{\epsilon}^{\rm sy}$ at the dimensionless 
synchrotron peak energy $\epsilon_{\rm sy}$ is approximately given by

\begin{equation}
f_{\epsilon}^{\rm sy} = (D \, B)^{7/2} \, {\pi \, c \, \sigma_{\rm T} 
\over 288 \, d_L^2} \, \left( [1 + z] \, \epsilon_{\rm sy} \, B_{\rm cr} 
\right)^{1/2} \, { p - 2 \over e_B \, m_e c^2}
\label{f_sy}
\end{equation}
where $B_{\rm cr} = 4.414 \times 10^{13}$~G. Note that the electron
spectrum normalization used to derive Eq. \ref{f_sy} is based on the
synchrotron spectrum above the synchrotron peak, where the underlying 
electron spectrum always has an index of $p \ge 3$. Eq. \ref{f_sy} 
yields a magnetic-field estimate of

\begin{equation}
B_{e_B} = 9 \, D_1^{-1} \left( {d_{27}^4 \, f_{-10}^2 \, e_B^2 \over
[1 + z]^4 \, \epsilon_{\rm sy, -6} \, R_{15}^6 \, [p - 2]} \right)^{1/7}
\; {\rm G},
\label{B_equipartition}
\end{equation}
where $f_{-10} = f_{\epsilon}^{\rm sy}$/(10$^{-10}$~ergs~cm$^{-2}$~s$^{-1}$),
$\epsilon_{\rm sy, -6} = \epsilon_{\rm sy}/10^{-6}$, and $R_{15} = R_B
/ (10^{15}$~cm).
With $d_{27} = 0.87$, $f_{-10} = 1$, $\epsilon_{\rm sy, -6} = 4$, $R_{15}
= 2$, and $p = 3.4$, this yields 

\begin{equation}
B_{e_B} = 3.6 \, D_1^{-1} \, e_B^{2/7} \; {\rm G}.
\label{B_eB}
\end{equation}

This is in excellent agreement with the estimate from the X-ray -- optical
delay, if the Doppler factor is slightly larger than 10 and/or the magnetic-field
equipartition parameter is slightly less than 1. We thus conclude that a
magnetic field of $B \sim 2 \, e_B^{2/7}$~G might be a realistic value for 
BL~Lacertae. We point out that the estimates in Eq. \ref{B_delay} and 
\ref{B_equipartition} should be valid for any model which represents the 
low-energy component of the blazar SED as synchrotron emission from relativistic 
electrons, which is the case for virtually all variations of leptonic and hadronic 
jet models.

\subsection{\label{other_parameters}Other relevant parameters}

Based on the magnetic-field estimate of 1.5 -- 2~G, the approximate location 
of the synchrotron peak of the SEDs of BL~Lacertae at $\nu_{\rm sy} \sim
10^{14}$~Hz allows us to estimate that the electron energy distribution in 
the synchrotron emitting region should have a peak at $\langle\gamma\rangle 
\sim 1.4 \times 10^3 \, D_1^{-1/2}$. The location of the synchrotron cutoff in
the quiescent state at $\nu_{\rm sy, co}^{\rm qu} \lesssim 10^{17}$~Hz then 
yields a maximum electron energy in the quiescent state of $\gamma_2^{\rm qu} 
\lesssim 4 \times 10^4 \, D_1^{-1/2}$, while the synchrotron cutoff in the
flaring state at $\nu_{\rm sy, co}^{\rm fl} \sim 2.4 \times 10^{18}$~Hz yields
$\gamma_2^{\rm fl} \sim 2 \times 10^5 \, D_1^{-1/2}$.

The superluminal-motion measurements mentioned in the introduction place a
lower limit on the bulk Lorentz factor $\Gamma \gtrsim 8$, and we expect that
the Doppler boosting factor $D$ is of the same order. Since, unfortunately, we
only have upper limits on the VHE $\gamma$-ray flux during our campaign, and
no measurements in the MeV --- GeV regime, no independent estimate from
$\gamma\gamma$ opacity constraints can be derived. However, such an estimate
was possible for the July 1997 $\gamma$-ray outburst and yielded $D \gtrsim
1.4$ \citep{bb00}, which is a much weaker constraint than derived from the 
superluminal motion observations. From the optical and X-ray variability time 
scale, we find an upper limit on the source size of $R_B \lesssim 1.6 \times 
10^{15} \, D_1$~cm. 

If the electrons in the jet are efficiently emitting most of their co-moving
kinetic energy before escaping the emission region (fast cooling regime),
then the kinetic luminosity of the leptonic component of the jet would have
to be $L_j^e \gtrsim 4 \pi \, d_L^2 \, (\nu F_{\nu})^{\rm pk} / D^4 \sim
10^{41} \, D_1^{-4}$~ergs~s$^{-1}$. If the electrons are in the slow-cooling
regime (i.e. they maintain a substantial fraction of their energy before
escaping the emitting region) and/or the jet has a substantial baryon
load \citep[for a recent discussion see, e.g.,][]{sm00}, the kinetic energy 
of the jet would have to be accordingly larger.

In both leptonic and hadronic blazar models, one needs an estimate of
the energy density in the external photon field. In the case of leptonic 
models, this determines the external-Compton processes, in hadronic 
models, p$\gamma$ processes on external photons depend on this quantity. 
For this purpose, an estimate of the average distance of the BLR from 
the central engine is needed, which can be achieved in the following 
way. The most recent determination of the mass of the central black 
hole in BL~Lacertae can be found in \cite{wu02}. They find a value 
of $M_{\rm BH}= 1.7 \times 10^8 \, M_{\odot}$. Then, if the width 
of the emission lines measured by \cite{ver95}, \cite{cor96}, and 
\cite{cor00} is interpreted as due to Keplerian motion of the BLR 
material around the central black hole, we find an estimate of the 
average distance of the line producing material of ${\bar r}_{\rm BLR} 
\sim 4.7 \times 10^{-2}$~pc. 

All of these estimates are model independent and provide a generic
framework for all relativistic jet models (in particular, leptonic
as well as hadronic models) aimed at reproducing the broadband SEDs 
and variability of BL~Lacertae during our campaign.

\section{\label{summary}Summary}

We have presented the observational results of an extensive multiwavelength
monitoring campaign on BL~Lacertae in the second half of 2000. The campaign
consisted of simultaneous or quasi-simultaneous observations at radio, optical,
and X-rays frequencies. Also, a simultaneous upper limit at $> 0.7$~TeV was
obtained with the HEGRA atmospheric \v Cerenkov telescope facility. We have
presented light curves, spectral variability characteristics, and broadband 
SEDs of BL~Lacertae during our observations. We have also looked for 
cross-correlations and time lags between different frequency bands as 
well as between narrow energy bands within the same frequency bands. 

The WEBT optical campaign achieved an unprecedented time coverage, virtually 
continuous over several 10 -- 20~hour segments. It revealed intraday variability 
on time scales of $\sim 1.5$~hours and evidence for spectral hardening associated 
with increasing optical flux. The multiwavelength campaign included two $\sim 25$ 
-- 30~ksec pointings with the {\it Beppo}SAX satellite. 

During the campaign, BL~Lacertae underwent a major transition from a rather 
quiescent state prior to September 2000, to a long-lasting flaring state 
throughout the rest of the year. This was also evident in the X-ray activity 
of the source. The {\it Beppo}SAX observations on July 26/27 revealed a rather 
low X-ray flux and a hard spectrum, while a BeppoSAX pointing on Oct. 31 -- 
Nov. 2, 2000, indicated significant variability on time scales of $\lesssim$~a 
few hours, and provided evidence for the synchrotron spectrum extending out to 
$\sim 10$~keV during that time.

Details of the data analyses as well as results pertaining specifically to the
optical and X-ray observations have been published in two previous papers on this
campaign \citep{villata02,ravasio03}. The new results presented in this paper 
for the first time include:

\begin{itemize}

\item We found a weak, low-significance indication of a delay of the 14.5~GHz 
light curve behind the 22~GHz light curve of $\sim 15$~d. No significant delays 
between the 14.5~GHz light curve and the lower-frequency light curves on the 
time scales covered by our campaign ($\lesssim 1/2$~yr) were detected.

\item We found that the optical intraday variability during the first BeppoSAX 
observation on July 26/27 seems to trace the soft X-ray variability with a time 
delay of $\sim 4$ -- 5~hr. If this delay is real and the result of synchrotron 
cooling of ultrarelativistic electrons, we can derive a magnetic field estimate 
of $B_{\rm delay, RX} = 1.6 \, D_1^{-1/3} \, (1 + k)^{-2/3}$~G. 

\item An additional, model-independent estimate of the magnetic field in the 
BL~Lacertae jet system from equipartition arguments yielded $B_{e_B} = 3.6 \, 
D_1^{-1} \, e_B^{2/7} \; {\rm G}$. 

\item We suggest a new method to estimate the magnetic field from the width
of the autocorrelation function of light curves at different photon energies, 
assuming that differences in the ACF widths are a measure of the energy-dependent 
radiative cooling time scale. In the case of the data available from our campaign,
this energy dependence is poorly constrained. However, taking the best-fit parameters 
at face value, we demonstrate our new method and find magnetic field estimates 
which are somewhat lower than inferred from the optical -- X-ray delay and from 
the equipartition argument.

\item We investigated the correlation between X-ray spectral hardness and intensity.
The {\it RXTE}~PCA data show a general trend of spectral softening during the highest
flux states on the time scale of several days sampled by those observations. For the 
second {\it Beppo}SAX observation, we could isolate individual short-term flares of
a few hours, and plot hardness-intensity diagrams over those individual flares. At
soft X-rays, we confirm the general trend of spectral softening during flares, while
the medium-energy X-rays show the opposite trend. In addition, some (though not all)
short-term flares show weak evidence for X-ray spectral hysteresis. 

\end{itemize}

Our campaign has revealed an extremely rich phenomenology of X-ray spectral
variability features which provide great potential for a deeper understanding
of the nature and energetics of the jets of low-frequency peaked and intermediate 
BL~Lac objects. However, the detection of these spectral variability phenomena 
were at (or even beyond) the limits of the capabilities of the {\it Beppo}SAX 
instruments. We strongly encourage future observations with the new generation 
of X-ray telescopes, in particular {\it Chandra} and {\it XMM-Newton}, to provide 
a more reliable and detailed study of these X-ray spectral variability phenomena.

Detailed modeling of the SEDs and variability properties of BL~Lacertae measured 
during this campaign will be presented in a companion paper (B\"ottcher \& Reimer
2003, in preparation).

\acknowledgments
We thank the anonymous referee for a very quick review with very useful
comments, which have helped to clarify and improve the paper. 
We also thank Dr. A. Reimer for careful reading of manuscript and very 
useful suggestions. 
A. P. Marscher, S. G. Jorstad, and M. F. Aller were supported in part by
NASA through grant NAG5-11811.
The work of M. Ravasio and G. Tagliaferri was supported by the Italian Ministry
for University and Research. 
The work of M. Villata and C. M. Raiteri was supported by the Italian Space
Agency (ASI) under contract CNR-ASI 1/R/073/02.
The European Institutes belonging to the ENIGMA collaboration acknowledge EC 
funding under contract HPRN-CT-2002-00321.
The St.-Petersburg group was supported by Federal Programs ``Astronomy"
(grant 40.022.1.1.1001) and ``Integration" (grant B0029).
A. Sadun and M. Kelly would like to thank the faculty and staff at UC Boulder's
Sommers-Bausch Observatory for the generous opportunity to use their facilities.

\newpage

\newpage

\newpage

\newpage

\newpage

\newpage

\newpage

\newpage

\newpage

\newpage

\newpage

\newpage

\newpage


\newpage

\begin{deluxetable}{cccccccc}
\tabletypesize{\scriptsize}
\tablecaption{Summary of X-ray ({\it RXTE} PCA and {\it BeppoSAX} MECS) Observations 
and spectral analysis results for a single-power-law model with fixed $N_H = 2.5 \times 
10^{21}$~cm$^{-2}$. The spectral analysis results for the PCA have been obtained for a 
fitted energy range (3 -- 15)~keV, the {\it BeppoSAX} MECS results have been obtained 
through MECS spectral analysis over the range (1 -- 10)~keV \citep{ravasio03}. 
$\alpha$ is the energy spectral index.}
\tablewidth{0pt}
\tablehead{
\colhead{Instrument} & \colhead{Start time} & \colhead{End Time} & \colhead{Duration} & 
\colhead{$\alpha$} & \colhead{$F_{\rm 1 \, keV}$} & 
\colhead{$F_{\rm 2 - 10 \, keV}$}& 
\colhead{$\chi_{\nu}^2$/d.o.f.} \\
\colhead{} & \colhead{[UT]} & \colhead{[UT]} & \colhead{[s]} & 
\colhead{} & \colhead{[$\mu$Jy]} & 
\colhead{[$10^{-12}$~ergs~s$^{-1}$~s$^{-2}$]} & 
\colhead{} 
}
\startdata
PCA & July 26, 18:23:44 & July 26, 19:00:32 & 2208 & $0.9^{+0.7}_{-0.6}$ & 1.4 & 6.3 & 0.42 / 25 \\
MECS & July 26, 10:12:39 & July 27, 06:43:33 & 23309 & $0.8 \pm 0.1$ & 1.18 & 5.8 & 0.86 / 43 \\
\noalign{\hrule} \\
PCA & Nov. 2, 10:56:16 & Nov. 2, 11:29:36 & 2000 & $1.45^{+0.3}_{-0.25}$ & 10.3 & 19.7 & 0.45 / 25 \\
MECS & Oct 31, 20:46:55 & Nov. 2, 09:59:28 & 33661 & $1.6 \pm 0.05$ & 12.7 & 19.7 & 0.61 / 58 \\
\enddata
\label{x_obs}
\end{deluxetable}


\begin{thebibliography}{}

\bibitem[Aharonian et al., 2000]{aharonian00}Aharonian, F., et al., 2000,
A\&A, 353, 847

\bibitem[Aharonian et al., 2002]{aharonian02}Aharonian, F., et al., 2002,
A\&A, 384, L23




\bibitem[Bloom et al., 1997]{bloom97}Bloom, S. D., et al., 1997, 
ApJ, 490, L145

\bibitem[B\"ottcher \& Bloom, 2000]{bb00}B\"ottcher, M., \& Bloom, S. D., 2000, 
AJ, 119, 469

\bibitem[B\"ottcher, 2002]{boettcher02}B\"ottcher, M., 2002, in proc. 
``The Gamma-Ray Universe'', XXII Moriond Astrophysics Meeting, in press

\bibitem[B\"ottcher \& Chiang, 2002]{bc02}B\"ottcher, M., \& Chiang, J.,
2002, ApJ, 581, 127

\bibitem[B\"ottcher et al., 2002]{bmr02}B\"ottcher, M., Mukherjee, R., \&
Reimer, A., 2002, ApJ, 581, 143

\bibitem[Bregman et al., 1990]{bregman90}Bregman, J. N., et al., 1990, 
ApJ, 354, 574

\bibitem[Cardelli et al., 1989]{cardelli89}Cardelli, M. T., Clayton, G. C., \&
Mathis, J. S., 1989, ApJ, 345, 245

\bibitem[Carini et al., 1992]{carini92}Carini, M. T., Miller, H. R., Noble, J. C.,
\& Goodrich, B. D., 1992, AJ, 104, 15

\bibitem[Catanese et al., 1998]{catanese98}Catanese, M., et al., 1998, 
ApJ, 501, 616

\bibitem[Chadwick et al., 1999]{chadwick99}Chadwick, P. M., et al.,
1999, ApJ, 513, 161

\bibitem[Clements \& Carini, 2001]{cc01}Clements, S. D., \& Carini, M. T., 2001,
AJ, 121, 90

\bibitem[Corbett et al., 1996]{cor96}Corbett, E.  A. et al. 1996, MNRAS,
281, 737

\bibitem[Corbett et al., 2000]{cor00}Corbett, E. A., Robinson, A., Axon, D. J.,
\& Hough, J. H., 2000, MNRAS, 311, 485

\bibitem[de Jager \& Stecker, 2002]{djs02}de Jager, O., \& Stecker, F. W., 
2002, ApJ, 566, 738

\bibitem[Denn et al., 2000]{denn00}Denn, G. R., Mutel, L. R.,
\& Marscher, A. P., 2000, ApJS, 129, 61


\bibitem[Dermer \& Atoyan, 2002]{da02}Dermer, C. D., \& Atoyan, A. M., 2002,
ApJ, 568, L81



\bibitem[Edelson \& Krolik, 1988]{ec88}Edelson, R. A., \& Krolik, J. H.,
1988, ApJ, 333, 646



\bibitem[Georganopoulos \& Marscher, 1998]{gm98}Georganopoulos, M., \&
Marscher, A. P., 1998, ApJ, 506, L11


\bibitem[Hartman et al., 1999]{hartman99}Hartman, R. C., et al.,
1999, ApJS, 123, 79


\bibitem[Holder et al., 2003]{holder03}Holder, J., et al., 2003, 
ApJ, 583, L9

\bibitem[Horan et al., 2002]{horan02}Horan, D., et al., 2002, ApJ, 571, 753

\bibitem[Kataoka et al., 2000]{kataoka00}Kataoka, J., Takahashi, T.,
Makino, F., Inoue, S., Madejski, G. M., Tashiro, M., Urry, C. M.,
\& Kubo, H., 2000, ApJ, 528, 243

\bibitem[Kirk et al., 1998]{krm98}Kirk, J. G., Rieger, F. M., \&
Mastichiadis, A., 1998, A\&A, 333, 452

\bibitem[Krawczynski et al., 2002]{krawczynski02} Krawczynski, H., Coppi, P. S.,
\& Aharonian, F. A., 2002, MNRAS, 336, 721

\bibitem[Kusunose et al., 2000]{ktl00}Kusunose, M., Takahara, F., 
\& Li, H., 2000, ApJ, 536, 299

\bibitem[Li \& Kusunose, 2000]{lk00}Li, H., \& Kusunose, M., 2000, 
ApJ, 536, 729


\bibitem[Madejski et al., 1999]{madejski99}Madejski, G., et al.,
1999, ApJ, 521, 145

\bibitem[Mang et al., 2001]{mang01}Mang, O., et al., 2001, in proc. of 
the 27$^{\rm th}$ ICRC, 2658



\bibitem[Mattox et al., 2001]{mhr01}Mattox, J. R.,
Hartman, R. C., \& Reimer, O., 2001, ApJS, 135, 155

\bibitem[Miller et al., 1989]{miller89}Miller, H. R., Carini, M. T., \&
Goodrich, B. D., 1989, Nature, 337, 627

\bibitem[M\"ucke \& Protheroe, 2001]{muecke01}M\"ucke, A., \& Protheroe, R. J.,
2001, Astropart. Phys., 15, 121

\bibitem[M\"ucke et al., 2003]{muecke03}M\"ucke, A., Protheroe, R. J.,
Engel, R., Rachen, J. P., \& Stanev, T., 2003, Astropart. Phys., 
18, 593


\bibitem[Nesci et al., 1998]{nesci98}Nesci, R., Maesano, M., Massaro, E.,
Montagni, F., Tosti, G., \& Fiorucci, M., 1998, A\&A, 332, L1

\bibitem[Neshpor et al., 2001]{neshpor01}Neshpor, Yu. I., Chalenko, N. N.,
Stepanian, a. A., Kalekin, O. R., Jogolev, n. A., Fomin, V. P., \&
Shitov, V. G., 2001, Astronomy Reports, 45, 249



\bibitem[Punch et al., 1992]{punch92}Punch, M., et al., 1992, 
Nature, 358, 477

\bibitem[Quinn et al., 1996]{quinn96}Quinn, J., et al., ApJ, 456, L83

\bibitem[Raiteri et al., 2001]{raiteri01}Raiteri, C. M., et al., 2001,
A\&A, 377, 396

\bibitem[Ravasio et al., 2002]{ravasio02}Ravasio, M., et al.,
2002, A\&A, 383, 763

\bibitem[Ravasio et al., 2003]{ravasio03}Ravasio, M., Tagliaferri, G.,
Ghisellini, G., Tavecchio, F., B\"ottcher, M., \& Sikora, M., 2003,
A\&A, submitted

\bibitem[Ryter, 1996]{ryter96}Ryter, C. E., 1996, Astrophys. \& Space Sci.,
236, 285


\bibitem[Sambruna et al., 1999]{sambruna99} Sambruna, R., et al., 1999,
ApJ, 515, 140

\bibitem[Schlegel et al., 1998]{schlegel98}Schlegel, D. J., Finkbeiner, D. P.,
\& Davis, M., 1998, ApJ, 500, 525


\bibitem[Sikora et al., 1997]{sikora97}Sikora, M., Madejski, G., Moderski, R.,
\& Poutanen, J., 1997, ApJ, 484, 108

\bibitem[Sikora \& Madejski, 2000]{sm00}Sikora, M., \& Madejski, G., 2000,
ApJ, 534, 109

\bibitem[Speziali \& Natali, 1998]{sn98}Speziali, R., \& Natali, G., 1998, 
A\&A, 339, 382

\bibitem[Takahashi et al., 1996]{takahashi96}Takahashi, T., et al., 1996,
ApJ, 470, L89

\bibitem[Vermeulen et al., 1995]{ver95} Vermeulen et al. 1995, ApJ, 
452, 5

\bibitem[Villata et al., 2000]{villata00}Villata, M., et al., 2000,
A\&A, 363, 108

\bibitem[Villata et al., 2002]{villata02}Villata, M., et al., 2002,
A\&A, 390, 407

\bibitem[Villata et al., 2003]{villata03}Villata, M., et al., 2003,
in preparation

\bibitem[Weekes et al., 2002]{weekes02}Weekes, T. C., et al., 2002, 
Astropart. Phys., 17, 221

\bibitem[Wu \& Urry, 2002]{wu02}Wu, J. H., \& Urry, C. M., 2002, 
ApJ, 579, 530

\end{thebibliography}
\end{document}